\documentclass[final,5p,times,twocolumn,authoryear]{elsarticle}

\usepackage{graphicx}
\usepackage{txfonts}
\usepackage[colorlinks=true,linkcolor=blue,citecolor=blue, urlcolor=blue]{hyperref}
\usepackage[justification=centering]{caption}
\usepackage{physics}
\usepackage[separate-uncertainty=true, binary-units]{siunitx}
\usepackage{amsmath}
\usepackage{empheq}
\usepackage{enumitem}
\usepackage{subcaption}

\usepackage{amsmath}	
\usepackage{algorithm}
\usepackage{placeins}
\usepackage{algpseudocode}
\usepackage{caption}
\usepackage{subcaption}
\usepackage{float}
\usepackage{xcolor}

\usepackage{multirow}
\usepackage{booktabs}

\newcommand{\bipp}{\texttt{BIPP}~}
\newcommand{\BIPP}{\texttt{BIPP}~}
\newcommand{\wsclean}{\texttt{WSClean}}
\newcommand{\WSClean}{\texttt{WSClean}}

\newcommand{\CASA}{\texttt{CASA}}

\journal{Astronomy $\&$ Computing}

\begin{document}

\begin{frontmatter}

\title{BIPP: An efficient HPC implementation of the Bluebild algorithm for radio astronomy}

\author[epfl,scitas]{Emma Tolley}
\author[cscs]{Simon Frasch}
\author[scitas]{Etienne Orliac}
\author[epfl]{Shreyam Krishna}
\author[epfl]{Michele Bianco}
\author[epfl]{Sepand Kashani}
\author[wsu]{Paul Hurley}
\author[epfl]{Matthieu Simeoni}
\author[epfl]{Jean-Paul Kneib}

\affiliation[epfl]{organization={Institute of Physics, Laboratory of Astrophysics, École Polytechnique Fédérale de Lausanne (EPFL)},
            addressline={Observatoire de Sauverny}, 
            city={Versoix},
            postcode={1290}, 
            country={Switzerland}}

\affiliation[scitas]{organization={SCITAS, École Polytechnique Fédérale de Lausanne (EPFL)},
            city={Lausanne},
            postcode={1015}, 
            country={Switzerland}}

\affiliation[cscs]{organization={Swiss National Supercomputing Centre (CSCS)},
            country={Switzerland}}

\affiliation[wsu]{organization={Western Sydney University},
            city={Penrith},
            country={Australia}}

\begin{abstract}
The Bluebild algorithm is a new technique for image synthesis in radio astronomy { which decomposes the sky into distinct energy levels using functional principal component analysis. These levels can be linearly combined to construct a least-squares estimate of the radio sky, i.e. minimizing the residuals between measured and predicted visibilities. This approach is particularly useful for deconvolution-free imaging or for scientific applications that need to filter specific energy levels.}
We present an HPC implementation of the Bluebild algorithm for radio-interferometric imaging: Bluebild Imaging++ (\texttt{BIPP}). 
The library features interfaces to C++, C and Python and is designed with seamless GPU acceleration in mind. We evaluate the accuracy and performance of \texttt{BIPP} on simulated observations of the upcoming Square Kilometer Array Observatory and real data from the Low-Frequency Array (LOFAR) telescope. We find that \texttt{BIPP} offers accurate wide-field imaging and has { competitive execution time with respect to the interferometric imaging libraries \CASA~ and \WSClean~ for images with $\leq 10^6$ pixels}. Futhermore, due to the energy level decomposition, images produced with \BIPP can reveal information about faint and diffuse structures before any cleaning iterations. {  \texttt{BIPP} does not perform any regularization, but we suggest methods to integrate the output of \texttt{BIPP} with CLEAN.}  The source code of \texttt{BIPP} is publicly released.
\end{abstract}


\end{frontmatter}



\section{Introduction}
Radio astronomers are engaged in ambitious new projects to detect faster, fainter, and more distant astrophysical phenomena. 
The flagship project is the Square Kilometer Array Observatory (SKAO)\footnote{\href{https://www.skao.int/}{https://www.skao.int/}}~\citep{ska2009}, 
and is considered { to be one of} the major ``Big Data'' challenges of the next decade.
According to the SKA estimates~\citep{skasdp}, the SKAO Science Data Processor (SDP) workflow will need to be able to deal with a data flow rate of around 1 TB/s at full capacity and will require a supercomputer of around 100 Pflops to calibrate and image the data.

 A major task of the SDP workflow will be \emph{image synthesis}: reconstructing an estimate of the radio sky $I$ using finite measurements of visibility space $V \in  \mathbb{C}^{L \times L}$. Image synthesis involves Fourier-transforming the calibrated visibilities into the ``dirty'' image (also known as the backprojected image). After imaging, deconvolution corrects the resulting images for the incomplete sampling of the Fourier plane. This step is typically performed using the CLEAN family of algorithms~\citep{clean1974}, which use a point-source model and iterative deconvolution to extract a final ``clean'' image of the sky.

Imaging is one of the most computationally demanding steps in the data processing pipeline, requiring significant memory and computing power. This complexity arises from operations like gridding, the convolutional resampling of observed data onto a regular grid, and fast Fourier transforms (FFT; \cite{FFTW97,FFTW05}) to transition between visibility and image space. The computational demands increase for observations of large fields of view (FoVs) where curvature effects cannot be neglected. Various algorithms have been developed to leverage the log-linear complexity of the 2D FFT,
{ such as the w-stacking algorithm~\mbox{\citep{offringa-wsclean-2014}} 
used by \texttt{WSClean}, the the w-projection technique~\mbox{\citep{wproj2008}} used by
\texttt{CASA}~\mbox{\citep{casa2022}}},
and faceting~\citep{faceting1992} used by \texttt{DDFacet}~\citep{ddfacet2018}. 
{ Many other techniques have been explored for the different steps of image synthesis}, for example using 3D NUFFTs~\citep{hvox2023},  optimization methods~\citep{usara2023} , or artificial intelligence~\citep{radionets2022}.

The Bluebild  algorithm~\citep{KashaniBluebild} { offers a new approach to image synthesis. Bluebild is a }method for reconstructing a least-squares consistent image of the sky based on the theory of sampling and interpolation operators~\citep{vetterli2014}. Image formation is formulated as a continuous inverse problem, and a key innovation of  Bluebild is to use  functional principal component analysis (fPCA) decomposition to simplify calculations of the sky image. Bluebild addresses the gridding, fast Fourier transform, and w-correction steps of imaging.

Image synthesis algorithms for next-generation radio interferometers must be able to handle the data sizes of the future SKAO. Their implementaions must be parallelizable, scalable, and portable.
In this paper, we provide an overview of the Bluebild algorithm and present our HPC implementation \texttt{BIPP}: Bluebild Imaging++. We show extensive validation results, comparing the output of \BIPP to \WSClean {\color{violet}~\citep{offringa-wsclean-2014,offringa-wsclean-2017}} and \CASA~\citep{casa2022}, and evaluate the performance of the \BIPP library. Finally, we explore the effect of the fPCA decomposition on image reconstruction and discuss scientific applications and future directions.



\section{The Bluebild Algorithm}

\subsection{ { Astronomical Measurement Equations}}

Electric fields emitted by astrophysical sources { can be} described by a continuous complex distribution  $E(\vec r,f,t)$,  where $\vec r$ is a (unit) direction vector denoting the sky coordinate. A radio antenna indexed by $p$ located at position { $\vec x_p \in \mathbb{R}^3$} will measure the phased sum of these electric fields {  over the entire sky as a voltage $v_p \in \mathbb{C}$ given by}~\citep{efieldradio}:
\begin{equation}
   v_p(f,t) ~= \int   E(\vec r, f, t) ~ g_p(\vec r, f, t) ~ \phi_p(\vec r, f) ~ d\Omega  ~+~ n_p(f,t)~,
   \label{eq:voltsandS}
\end{equation}
where $g_p(\vec r, f, t)$ is the directional antenna voltage response at a given frequency and time,
{ $\phi_p(\vec r,f) \equiv  \exp { \bigr(- \frac{2 \pi i f}{c}  \langle \vec x_p , \vec r\rangle \bigl)}$ }is the steering vector of the instrument~\citep{Simeoni2019}, also known as the phase delay term,
and $n_p(f,t)$ is { uncorrelated} additive receiver noise.
{  If we discretize $\vec r$ into $N_\text{sky}$ discrete coordinates indexed by $j$, we can write the discrete form of Eq.~\ref{eq:voltsandS}:
\begin{equation} \label{eq:discretev}
   v_{p}(f,t) = \sum_{j}^{N_\mathrm{sky}} E_j(f,t)~ g_{pj}(f,t) ~ e^{- \frac{2 \pi i f}{c}  \langle \vec x_{p}, \vec r_j \rangle} + n_p(f,t) .
\end{equation}
By introducing introduce a time- and frequency-dependent matrix known as the sampling operator
$\Psi^* \in \mathbb{C}^{N_A\times N_\mathrm{sky}}$ with elements defined as
\begin{equation}
   \Psi_{pj}^* = g_{pj}e^{\frac{2 \pi i f}{c} \langle \vec x_p, \vec r_j  \rangle},
\end{equation}
we can write Eq.~\ref{eq:discretev} concisely in matrix notation:
\begin{equation}
   \vec v = \Psi^* \vec E + \vec \eta,
\end{equation}
where $\vec v \in \mathbb{C}^{N_A}$ is the vector of voltages recorded over all $N_A$ antennas, $\vec E  \in \mathbb{C}^{N_\text{sky}}$ is the vector of electric fields across the discretized sky coordinates, and $\vec \eta \in \mathbb{C}^{N_A}$ is the vector of noise terms across all antennas. For conciseness we have dropped the explicit dependence on frequency and time, but note that $\Psi$, $\vec \eta$, $\vec E$, and $\vec v$ are all frequency- and time-dependent.

Interferometers measure visibilities, the voltage correlation between two antennas $p$ and $q$~\citep{rime1996,rime2009,rime,vanderVeen2013}. The voltage correlation matrix, also known as the visibility  matrix, is given by:
\begin{align}
   V &\equiv \mathbb{E}[\vec v \vec v^*] = \Psi^* B \Psi + \sigma^2 \mathbb{I},
\end{align}
where $\mathbb{I}$ is the identity matrix, the noise correlation matrix is  $\sigma^2 \mathbb{I} = \mathbb{E}[\vec \eta ~\vec \eta^*]$, and $B$ is the correlation matrix of the discretized electric field emission $\vec E$: $B = \mathbb{E}[\vec E \vec E^*]$. Because signals coming from different directions in the sky are uncorrelated~\citep{synthesisimaging1999}, $B$ is a diagonal matrix with diagonal elements given by $B_{jj} = |E_j|^2$. For concise notation in later derivations, we  define the sky brightness vector $\vec I$ as the diagonal of $B$:
\begin{equation}
    I_j \equiv B_{jj} = |E_j|^2.
\end{equation}

\subsection{The Inverse Problem}

Image synthesis in radio astronomy seeks to reconstruct an estimate of $B$ from discrete measurements of $V$:
\begin{equation}
    V = \Psi^*B\Psi + \sigma^2 \mathbb{I}.
\end{equation}
We can try to construct an estimate of $B$ called $\widetilde B$ which satisfies the \emph{least-squares} problem, i.e. minimizes the following expression:
\begin{equation} \label{eq:lsq}
    || \Psi^* \widetilde B \Psi - V ||^2,
\end{equation}
The solution to this is well known and takes the form:
\begin{equation}
   \widetilde B = \Psi G_\Psi^{-1} ~ V ~ G_\Psi^{-1} \Psi^*~,
   \label{eq:skyreco}
\end{equation}
where $G_\Psi \in \mathbb{C}^{N_A \times N_A}$ is the {\em Gram matrix}  of the instrument defined as $G_\Psi \equiv \Psi^* \Psi$}:
\begin{equation}
    (G_\Psi)_{pq} = \sum_j^{N_\text{sky}} \Psi_{pj}^* \Psi_{jq} =   \sum_j^{N_\text{sky}} g_{pj} g_{qj}^* e^{2 \pi i \langle \vec b_{pq}, \vec r_j \rangle} 
\end{equation}
where $\vec b_{pq}$ is the baseline vector between antennas $p$ and $q$: $\vec b_{pq} \equiv \frac{f}{c}(\vec x_p - \vec x_q)$.
In the case of calibrated omnidirectional antennas we can write $g_p(\vec r) = 1$, and the Gram matrix can be shown to have the following analytical closed form (at the continuous level)~\citep{KashaniBluebild}:
\begin{equation}
(G_\Psi)_{pq} = \mathrm{sinc}~\Bigl(2~ \Bigl|\Bigl|  \vec b_{pq}  \Bigr|\Bigr| \Bigr)~.
\label{eq:gram}
\end{equation}
Unfortunately, the Gram matrix can be  ill-conditioned~\citep{taylor_1978}, hence evaluating its inverse in Eq.~\ref{eq:skyreco} is prone to error.

\subsection{Functional Principal Component Analysis}

The Bluebild algorithm~\citep{KashaniBluebild} calculates $\widetilde {B}$ and thus $\widetilde {I}$ from Equation~\ref{eq:skyreco} by directly finding a decomposition of $\widetilde{B}$ in a compact orthogonal basis, namely
\begin{equation}
    \widetilde{B} 
    = \sum_{a} \lambda_{a} \epsilon_{a} \epsilon_{a}^{H} 
    = \sum_{a} \lambda_{a} \Psi \alpha_{a} \alpha_{a}^{H} \Psi^{*},
\end{equation}
where $\{(\lambda_{a}, \, \epsilon_{a} \equiv \Psi \alpha_{a})\} \in \mathbb{R} \times \mathbb{C}^{N_{\mathrm{sky}}}$ 
are eigenpairs of $\widetilde{B}$.
The relation $\epsilon_{a} = \Psi \alpha_{a}$, where $\alpha_a \in \mathbb{C}^{N_A}$ follows since $\epsilon_{a} \subset \text{span}(\Psi)$.
The parameters $\{(\lambda_{a}, \alpha_{a})\}$ can easily be inferred starting from the eigenvalue property:
\begin{equation*}
    \widetilde{B} \epsilon_{a} = \lambda_{a} \epsilon_{a}~.
\end{equation*}
Combining the above with the expression for $\widetilde{B}$ in Equation~\ref{eq:skyreco} and using  $\epsilon_{a} \equiv \Psi \alpha_{a}$ we obtain:
\begin{equation*}
    \bigl( \Psi G_\Psi^{-1} ~ V ~ G_\Psi^{-1} \Psi^* \bigr) ~ \Psi \alpha_a = \lambda_a \Psi \alpha_a
\end{equation*}
Using $G_\Psi = \Psi^*\Psi$, this reduces to:
\begin{equation}
    V  \alpha_a = \lambda_a G_\Psi \alpha_a.
    \label{eq:eigenvaleq}
\end{equation}
Thus the parameters $\{(\lambda_{a}, \alpha_{a})\}$ are obtained by solving the generalized eigenvalue problem $ V \alpha_a = \lambda_a G_{\Psi} \alpha_a$. This allows us to calculate the least-squares solution without inverting $G_{\Psi}$, giving: 
\begin{equation}
    \widetilde {I} = \sum_{a} \lambda_{a} |\epsilon_{a}|^2 = \sum_{a} \lambda_{a} |\Psi \alpha_{a}|^2~.
    \label{eq:eigenvaluedeco1}
\end{equation}
Recall that $V$ and $\Psi$ are time and frequency dependent, so $\widetilde{I}$ is an estimate of the instantanous narrow-band sky intensity. Thus the eigenvalue decomposition in Eq.~\ref{eq:eigenvaleq} { must be} repeated for each timestep $t$ and frequency band $f$. {  It can be shown that the time-integrated LSQ image $\sum_t \widetilde I(t)$ will also be a LSQ solution if the corresponding Gram matrix for the integrated operator is block-diagonal, which is true if $|| \vec x_p(t) - \vec x_q(t+\Delta t)|| \gg \lambda$ for all $p$ and $q$, with wavelength $\lambda = f/c$ and integration time $\Delta t$. For $\lambda = 6\text{m}$ and an observing latitude of $30^\circ$ this assumption holds for $\Delta t > 0.02 \text{s}$.}

\subsection{Standard Image Synthesis}
\label{subsub:ss_img_synt}
After obtaining our eigenpairs $(\lambda_a, \alpha_a)$,
we can reconstruct  $\epsilon_a$ by {  directly applying the complex conjugate of the sampling operator} $\Psi$:
\begin{equation}
 \epsilon_{aj} = \sum_p  \Psi_{jp} \alpha_{ap} = \sum_p g^*_{pj} ~e^{ -\frac{2 \pi i f}{c} \langle \vec x_{p},\vec r_j \rangle } \alpha_{ap}~.
 \label{eqn:standard_synthesis}
\end{equation}
Directly calculating this result via matrix multiplication is called {\em Standard Synthesis}.
Because $\vec b_{p}$ is comprised of the instantaneous antenna positions, $\Psi$ is a time- and frequency-dependent operator, and Eq.~\ref{eqn:standard_synthesis} must be evaluated at each timestep. 

\subsection{NUFFT Image Synthesis}\label{subsub:nuftt_img_synt}
We can improve Standard Synthesis by leveraging algorithms for {  non-uniform FFT~\citep[NUFFT;][]{LEE20051} of type-3, which maps from a non-uniform input domain to a non-uniform output domain~\citep{nudft}.}
The expression for our least-squares reconstructed sky can be expanded as
\begin{equation}
  \widetilde {I} = \sum_{a} \lambda_a | \epsilon_{a}|^2 = \Psi ~V' ~\Psi^*~,
 \label{eqn:synthesis}
\end{equation}
where $V'$ are the Gram-corrected visibilities { 
\begin{equation}
    V' = A ~ \Lambda~ A^H~,
\end{equation}
where the columns of matrix $A \in \mathbb{C}^{N_A \times N_A}$ are the ordered eigenvectors $\alpha_a$ and $\Lambda$ is a diagonal matrix with diagonal elements as the ordered eigenvalues $\lambda_a$.
Writing out $\Psi$ explicitly and assuming calibrated gains $g_p(\vec r) = 1$,  Equation~\ref{eqn:synthesis} can be expressed as a 
discrete Fourier transform:
\begin{equation}
\widetilde I_j = \sum_p^{N_A} \sum_q^{N_A} V'_{pq} ~ e^{-2 \pi i \langle \vec b_{pq},  \vec r_j \rangle}~.
\end{equation}
 We can consolidate the sum over separate baselines by choosing an appropriate index $n$ for the baselines created by antenna pairs $p$ and $q$:
\begin{equation}
\widetilde I_j = \sum_n^{N_A \times N_A} V'_n ~ e^{-2 \pi i \langle \vec b_n,  \vec r_j \rangle}~,
\label{eq:type3nufft}
\end{equation}
We have $N_A^2$ samples of the visibilities $V'_n$ at coordinates $\vec b_n$ in the $uvw$ plane, and map to $N_\text{sky}$ coordinates in the sky.  This can be evaulated with a type-3 NUFFT~\citep{hvox2023}, which performs the familiar gridding/degridding operations as defined in Section~\ref{sec:nufftsyn}. We call this alternative imaging strategy \emph{NUFFT Synthesis}.}

\subsection{Energy Levels \& Partitioning}
\label{subsec:energy_levels_and_partitioning}
A key aspect of the Bluebild algorithm is decomposing the sky into { distinct eigenvalue-eigenvector pairs $\{ (\lambda_a, \alpha_a)\}$ as} shown in Equation~\ref{eq:eigenvaluedeco1}. These levels can be manipulated via truncation, partitioning, or filtering to create different output images.

{\bf Truncation}: not every eigenvector $\alpha_a$ needs to be used for constructing the LSQ sky intensity estimate $\widetilde{I}$. As the eigenvectors with the smallest eigenvalues often correspond to noise, \texttt{BIPP} includes an option only to construct images using $N_\mathrm{eig} \leq N_A$ leading eigenvectors. $N_\mathrm{eig}$ can be set to the total number of eigenvalues $N_A$,  a custom value defined by the user, or estimated from a  given observation by determining the minimum number of leading eigenvectors that
account for a user-defined percentage of the total energy. Discarding the smallest-energy levels allows for efficient suppression of noise in the reconstructed image. { Alternatively, the largest eigenvectors can be discarded to remove the brightest point sources.}

{\bf Partitioning}: for the imaging process, the $\{(\lambda_{a}, \alpha_a)\}$  pairs can optionally be  partitioned or grouped {  into $N_L \leq N_A$ energy levels. A level consists of $n$ eigenpairs $i_0$ through $i_n$. In Standard Synthesis, this is defined by:
\begin{equation}
    \widetilde{I}_{i_0,i_n} = \sum_{a = i_0}^{a=i_n} \lambda_a |\epsilon_a|^2
\end{equation}
In NUFFT synthesis the levels are defined at the visibility level, minimizing the numbers of calls to the type-3 NUFFT:
 \begin{equation}
\widetilde I_{i_0,i_n} = \Psi ~ V'_{i_0,i_n} \Psi^*.
\end{equation}
$V'_{i_0,i_n}$ is constructed by setting eigenvalues which fall outside the interval to zero.
}
%
%
Level partitions can be automatically determined using $k$-means clustering on the eigenvalues $\{\lambda_{a}\}$, or defined as intervals by the user. Combining the $N_A$ eigenvectors into a smaller number of levels reduces the total number of calls to Standard Synthesis or NUFFT Synthesis during imaging. Examples of partitioned images output by \BIPP are shown in Section~\ref{sec:SA}.

Finally, eigenvalues and eigenvectors can be {  using user-defined {\bf filters} to create different images with different weights}. One  possible filter is {\em Least-Squares (LSQ)},
which combines the eigenvectors $\epsilon_{a}$ after re-weighting them at their true scale $\lambda_{a}$ as in Equation~\ref{eq:eigenvaluedeco1}, thus producing an image minimizing the least-squares optimization problem. Alternatively, a { \em Standardized} image can be constructed by performing a uniform sum across eigenvectors $\{\epsilon_{a}\}$, effectively normalizing the flux across all eigenimages.

\subsection{Beamforming}\label{sec:beamforming}
For certain telescopes, the $N_A$ antenna { voltages $\vec v$} are not available directly but are instead beamformed together into $K$ beams using a weighting matrix $W \in \mathbb{C}^{N_A\times K}$~\citep{beamforming}. The beamformed { voltages $\vec v^W \in \mathbb{C}^K$} are then given by:
\begin{equation}
    \vec v^W_k = W \vec v.
\end{equation}
We can define a modified sampling operator, which includes beamforming $\Phi^* = W^H \Psi^*$.
The beamformed visibilities $V^W \in  \mathbb{C}^{K \times K}$ are thus given by:
\begin{equation}
    V^W = \Phi^* ~B~ \Phi =W^H ~\Psi^* ~B~ \Psi ~W~,
\end{equation}
and with { this new sampling operator} the Gram matrix $G_\Psi^W \in  \mathbb{C}^{K \times K}$ becomes:
\begin{equation}
  G_\Psi^W = \Phi^* ~ \Phi =W^H ~\Psi^* ~ \Psi ~W~.
\end{equation}
Beamforming can easily be accommodated in Standard Synthesis and NUFFT Synthesis with the redefined sampling operator $\Phi^*$.

{ 
\subsection{Comparison with CLEAN}\label{sec:vsclean}

The CLEAN algorithm~\citep{clean1974} and its variants~\citep{1980clark,Cornwell_2008} use an iterative method to construct an estimate of $B$~\citep{vandertol-2018}. \emph{Back-projection} is used to construct an initial estimate $\hat B$:
\begin{equation}
    \hat B = \Psi V \Psi^*.
\end{equation}
We can write out $\hat B$ explicitly:
\begin{equation}\label{eq:cleanlike}
   \hat B_{jj} = \hat I_{j} =  \sum_p^{N_A} \sum_q^{N_A} \Psi_{pj} V_{pq} \Psi_{qj}^* =
 \sum_{s}^{N_A \times N_A} V_{n} ~ e^{ -2 \pi i   \langle \vec b_{n},\vec r_j \rangle},
\end{equation}
where $\hat I_{j}$ is our estimate of the sky intensity at each sky coordinate and again we assume calibrated gains $g_p(\vec r) = 1$. Note that this looks very similar to Eq.~\ref{eq:type3nufft}; the definition of the input and output coordinates are the same. 
If we choose a regular sampling of the discrete sky coordinates $r_j$ in $lmn$ coordinates while assuming negligible curvature of the sky ($\sqrt{1-l^2 - m^2} \simeq 0$) and express the antenna baselines $\vec b_{n}$ into the familiar frequency-dependent $uvw$ frame we obtain:
 \begin{equation}
    \hat I[l,m] = \sum_u \sum_v V[u,v]e^{- 2 \pi i (ul + vm)} ~,
\end{equation}
where $V[u,v]$ is the map of samples in the $uv$ plane with sample positions given by the baseline vectors and sample strength given by $V$. Thus $\hat I$ is the familiar \emph{dirty image}
 often created as an intermediate step { before deconvolution} by state-of-the-art radio interferometry imaging software such as  \texttt{WSClean}. Deconvolution removes an effect called the point spread function (PSF) which is determined by the baseline geometry $\vec b_n$:
\begin{equation}
I^\text{PSF}_j = \sum_n^{N_A \times N_A} e^{-2 \pi i \langle \vec b_n,  \vec r_j \rangle}~,
\end{equation}
The back-projected image can be represented as the convolution of the true sky image $I^\text{sky}$ with the PSF:
\begin{equation}
\hat I = I^\text{PSF} * I^\text{sky}
\end{equation}

 The target ``clean'' image is constructed by iteratively adding components to the initially empty model image $B_k \in \mathbb{R}^{N_{sky}}$, where components are selected using the residual $R_k \in \mathbb{R}^{N_{sky}}$. For example, if we use a dirac delta $\delta_j$ located at sky coordinate $j$ for our component, then 
 \begin{align*}
R_k &= \hat B - \Psi^* B_{k-1} \Psi \\
B_k &= B_{k-1} + \alpha \delta_j, \quad \text{where}~ j = \underset{j}{\operatorname{argmax}} ~ R_k[j] 
 \end{align*}
where $\alpha$ is some multiplicative gain factor. The iterations are halted
 when the current iterate is considered good enough, for example when the residual noise is smaller than some pre-defined threshold or after a fixed number of iterations. With enough iterations $B_{k \rightarrow \infty}$ will satisfy the last-squares minimization problem of Eq.~\ref{eq:lsq}, with additional implicit regularization provided by the component model.
%

The estimate from Bluebild $\widetilde B$ gives us a least-squares solution without any cleaning iterations.  This makes $\widetilde B$ a useful replacement for $\hat B$ for deconvolution-free imaging, which is foreseen for DSA-2000~\citep{2019dsa2000}. However, unlike $B_k$, $\widetilde B$ is constructed without any regularization terms, and still includes the effect of the PSF. This can easily be seen by comparing Equations~\ref{eq:type3nufft} and \ref{eq:cleanlike}: both expressions map $N_A^2$ samples at coordinates $\vec b_n$ in the $uvw$ plane to $N_\text{sky}$ coordinates in the sky. The discrete $\vec b_n$ sample coordinates create the PSF, but the sample values themselves change. Bluebild's construction of $\widetilde B$ use the Gram-corrected visibilities used by Bluebild $V'$, and the back-projection $\hat B$ uses the instrument visibilities $V=G_\Psi V'$, from the generalized eigenvalue problem.

Note that if we approximate the Gram matrix as the identity matrix $G_\Psi = \mathbb{I}$ then the back-projection and Bluebild least-squares estimate converge: $\hat B = \widetilde B$ and $\hat I = \widetilde I$.}
 Many modern interferometers are implicitly designed such that the off-diagonal terms of the Gram matrix are be small, and thus { $\hat I$ is a reasonable approximation of the least-squares solution $\widetilde I$}. However, for instruments with redundant baselines or low frequencies, the off-diagonal terms are larger, and the Gram matrix has a stronger effect,  as discussed in~\ref{sec:appGram}.

{ 
\subsection{Complexity analysis}\label{sec:complexity}

For a given timestep and frequency band, the eigenvalue decomposition can be computed in $\mathcal{O}(N_A^3)$ operations. The computational cost of the decomposition does not contribute significantly to the overall computation time because $N_A$ remains small; SKA-Mid in South Africa is expected to have $N_A = 197$ dishes, and SKA-Low in Australia is expected to have $N_A = 512$ stations. Compared to the imaging steps of the algorithm the execution time for the eigenvalue decomposition is small, as shown in Section~\ref{sec:perf}.
For an observation with $N_A$ dishes or stations, $T$ timesteps, and $F$ frequency bands, the computation time of the eigenvalue decomposition scales as $\mathcal{O}(T~F~N_A^3)$.

For Standard Synthesis, the $N_\text{A} \times N_\text{sky}$ terms of $ \epsilon_{aj}$ must be evaluated with a sum over $N_A$ elements, the same computational cost as a direct Fourier transform. The computational complexity scales as $\mathcal{O}(T~F~N_\text{sky}~N_A^2)$. Unlike $N_A$, $N_\text{sky}$ often has extremely large values.

The algorithmic complexity of the 3D type-3 NUFFT is~\citep{hvox2023}:
\begin{equation}
    \mathcal{O}\bigl(( N_\text{in} + N_\text{out})|\log \epsilon|^3  +  N_\text{out} + N_\text{mesh}( \log N_\text{mesh}+ 1)\bigr)~,
\end{equation}
where $N_\text{in}$ is the number of input points, $N_\text{out}$ is the number of output points, $\epsilon$ is the user-requested accuracy parameter, and $N_\text{mesh}$ is the uniform sampling which depends on the distribution of the input and output points.
For NUFFT Synthesis we can stack our visibility samples across timesteps to obtain $N_\text{in} \propto N_A^2 T$ input samples, and similarly stack our $N_A$ eigenvectors across $N_L \leq N_A$ levels.
Then, using the type-3 NUFFT for $F$ frequency channels has a leading order computational complexity of:
\begin{equation}
    \mathcal{O}\Bigl(FN_L \bigl(( N_A^2 T + N_\text{pix})|\log \epsilon|^3  +  N_\text{sky} + N_\text{mesh}( \log N_\text{mesh}+ 1)\bigr)\Bigr)~.
\end{equation}
Thus the the $T$ and $N_\text{sky}$ scaling is decoupled, improving the performance of NUFFT Synthesis compared to Standard Synthesis.

We can furthermore reduce the dependence on $F$ by stacking different frequency samples into the input points to the type-3 NUFFT, thus performing multi-frequency-synthesis. This functionality is not included in the current implementation of Bluebild, but is planned for a future release. 

}

\subsection{Previous implementations}

A Python implementation of Bluebild currently exists on GitHub,\footnote{\href{https://github.com/imagingofthings/pypeline}{https://github.com/imagingofthings/pypeline}} which includes extra software modules for reading in astronomical data, creating simulated data, interpolating output from the sphere to the plane, and writing output to the FITS file format~\citep{FITS}.

\vspace{-1em}
\section{Bluebild Imaging++}

Bluebild Imaging++ (\BIPP\hspace{-0.4em}) is an HPC implementation of Bluebild based on the Python implementation by \cite{KashaniBluebild}, where the core Bluebild algorithm has been rewritten in C++.
The \BIPP library features interfaces to C++, C and Python and is designed with seamless GPU acceleration in mind. An additional Python module offers data preprocessing functionality for common input formats used in radio astronomy. The initial release of \BIPP targets non-distributed computations only, but MPI support is planned for a later stage. The source code is publicly available on GitHub\footnote{\href{https://github.com/epfl-radio-astro/bipp}{https://github.com/epfl-radio-astro/bipp}}.

\subsection{Interface}

The interface to \BIPP is structured around two types. A \textit{Context}, which holds any reusable resources such as memory for either CPU or GPU processing, and an imaging handle called \textit{StandardSynthesis} or \textit{NUFFTSynthesis}, depending on which Bluebild image synthesis strategy is used.

An imaging handle can iteratively collect input data at each time step for a given set of filters and partitioning applied to eigenvalues. Through shared use of a \textit{Context}, multiple imaging handles can share resources to reduce overall consumption. When using a GPU as a processing unit, most input data can be located in the host or device memory. \BIPP will automatically transfer any data to device memory or vice versa as required. When using CPU processing, \BIPP can utilize any LAPACK and BLAS compatible library for solving the eigenproblem and linear algebra operations.

For computing the NUFFT, \texttt{FINUFFT}~\citep{finufft} is used. On GPU, \BIPP relies on CUDA for Nvidia hardware and HIP for AMD hardware. With either programming framework, the corresponding BLAS and LAPACK libraries cuBLAS, cuSOLVER, hipBLAS or MAGMA~\citep{magma} are utilized, in addition to the cuFINUFFT library, for which a special fork with the required transform type and HIP support is available\footnote{\href{https://github.com/AdhocMan/cufinufft/tree/t3\_d3}{https://github.com/AdhocMan/cufinufft/tree/t3\_d3}}.

\vspace{-0.5em}
\subsection{Input data}
\begin{table}
\begin{tabular}{lcl}
$N_\mathrm{eig}$ & : & Number of requested eigenvalues. \\
$N_f$ &:& Number of filters.    \\
$S \in \mathbb{C}^{K \times K}$ & : & Visibility matrix. \\
$W \in \mathbb{C}^{N_A \times K}$ & : & Beamforming matrix. \\
$P_\mathrm{icrs} \in \mathbb{R}^{N_A \times 3}$ & : & Stacked antenna positions in ICRS coords. \\
$P_{uvw} \in \mathbb{R}^{N_A^2 \times 3}$ & : & Stacked UVW coordinates. \\
$X_\mathrm{pix} \in \mathbb{R}^{N_\mathrm{pixel} \times 3}$ & : & Stacked image pixel coordinates. \\
$I \in \mathbb{R}^{N_i \times 2}$ & : & Stacked eigenvalue partition intervals \\
& & of form [min, max]. \\
\end{tabular}
\caption{Input variables to the \texttt{BIPP} imaging handle. { $K$ is equivalent to $N_A$ if beamforming is not used.}}
\label{tab:input}
\end{table}

\texttt{BIPP} uses a single function to process all data from a given single time step to enable better usage of GPU acceleration and overall interface simplicity.
In the initialization stage, an imaging handle requires the coordinates for each pixel, the set of filters to apply to the eigenvectors, and the number of expected eigenvalue partitions in the form of intervals. 
The \textit{collect} function expects the input shown in Table~\ref{tab:input} in column-major memory layout for a given number of antennas $N_A$, beams $K$, and eigenvalue interval partitions $N_i$.
Aside from the user-defined $N_\mathrm{eig}$ and $I$, all of these quantities can be extracted from real or simulated observations. \texttt{BIPP} is configured to read these quantities from standard CASA MeasurementSet~\citep{casams2000} files.

\vspace{-0.5em}
\subsection{Standard Synthesis}
Our implementation of the Standard Synthesis algorithm described in Section~\ref{subsub:ss_img_synt} is described in Algorithm \ref{alg:ss}. It uses a custom kernel for computing Eq. \ref{eqn:standard_synthesis} for each pixel, interval and filter, such that the number of complex exponential evaluations is minimized as shown in Algorithm \ref{alg:gemmexp}.

\begin{algorithm}[h]
\caption{Standard Synthesis}\label{alg:ss}
\begin{algorithmic}[1]
\Procedure{collect}{$N_\mathrm{eig}, W, S, I, P_\mathrm{icrs}, X_\mathrm{pix}$}
  \State $G \gets $ Compute Gram matrix $(W, P_\mathrm{icrs})$
  \State $\alpha, \lambda \gets $ Generalized eigenvalue decomposition $(S, G)$
  \State $\alpha, \lambda \gets $ Select $N_\mathrm{eig}$ largest eigenvalues ($\alpha, \lambda$)
  \State $E \gets \Call{gemmexp}{\alpha, W, P_\mathrm{icrs}, X_\mathrm{pix}}$
  \For{$n_f\in\{1,...,N_f\}$ }
  \State $\lambda_f \gets$ Apply filter ($n_f, \lambda$)
  \For{$n_i\in\{1,...,N_i\}$}
     \State $K \gets $ Indices of eigenvalues in interval $I^{(n_i)}$
     \State $B^{(n_i, n_f)} \gets B^{(n_i, n_f)} + \sum_{i \in K} d_f^{(i)} E^{(i)}$
  \EndFor
  \EndFor
  
  \State Store $B$
\EndProcedure
\end{algorithmic}
\end{algorithm}

\begin{algorithm}[h]
\caption{GEMMEXP kernel}\label{alg:gemmexp}
\begin{algorithmic}[1]
\Procedure{gemmexp}{$\lambda, \alpha, W, P_\mathrm{icrs}, X_\mathrm{pix}$}
  \State $U \gets WA$
  \For{$n_p\in\{1,...,N_\mathrm{pixel}\}$} 
  \State $R \gets 0$
  \For{$n_a\in\{1,...,N_A\}$} 
  \State $p \gets e^{i\frac{2 \pi}{\lambda} (P_\mathrm{icrs}^{(n_a)} \cdot X_\mathrm{pix}^{(n_p)})}$
  \For{$n_e\in\{1,...,N_\mathrm{eig}\}$}
  \State $R^{(n_e)} \gets R^{(n_e)} + pU^{(n_a, n_e)}$  
  \EndFor
  \EndFor
  \For{$n_e\in\{1,...,N_\mathrm{eig}\}$}
  \State $E^{(n_e, n_p)} \gets E^{(n_e, n_p)} + {\lVert R^{(n_e)} \rVert}^2$
  \EndFor
  \EndFor
  \State Return $E$
\EndProcedure
\end{algorithmic}
\end{algorithm}

\subsection{NUFFT Synthesis}\label{sec:nufftsyn}
We also implement the NUFFT Synthesis algorithm described in Section~\ref{subsub:nuftt_img_synt}.
However, direct evaluation of the $N$ complex terms of equation~\ref{eq:type3nufft} would involve
computing exponential sums that naively require $O(N_\mathrm{baseline} N_\mathrm{pixel} )$ effort.
Instead, the \texttt{FINUFFT} library~\citep{finufft} approximates the coefficients to within user-specified relative tolerance $\epsilon$, in close to linear time in $N_\mathrm{baseline}$ and $N_\mathrm{pixel}$. As neither the antenna baseline coordinates nor $3D$ sky pixel coordinates are uniformly distributed, we use the type-3 (nonuniform to nonuniform) NUFFT.

The first step is to rescale the 3D coordinates $b_n$ to lie within $[-\pi, \pi)^3$ with dilation factors $\gamma_i$. We then write equation ~\ref{eq:type3nufft} as:
\begin{equation}
\widetilde B_\mathrm{pix} = \sum_n^N V'_n e^{i x_\mathrm{pix}' \cdot b_n'}~,
\end{equation}
where $x_\mathrm{pix}'^{(i)} = \gamma_ix_\mathrm{pix}^{(i)} $ and  $b_n'^{(i)} = \gamma_ib_n^{(i)} $. These rescaled baseline coordinates are then spread onto a fine regular grid $b_x$ using a periodized kernel such that
\begin{equation}
b_x = \sum^N_{n} V_n \phi
\bigl(
x_1 h_1 - b'^{(1)}_n,
~ x_2 h_2  - b'^{(2)}_n,
~ x_3 h_2  - b'^{(3)}_n
\bigr)~,
\end{equation}
where $h_i$ is the fine grid spacing and $\phi$ is the normalized and periodicized ``exponential
of semicircle'' kernel from FINUFFT. The user-requested tolerance $\epsilon$ sets the kernel width and sampling.
These $b_x$ can be treated as Fourier series coefficients, and we can evaluate this series at rescaled target points
using the type-2 NUFFT (uniform to nonuniform):
\begin{equation}
b_\mathrm{pix}  = \sum_{x} b_x e^{ix\cdot \bigl( h_i \gamma_i x_\mathrm{pix}^{(i)} \bigr) }~.
\end{equation}
Lastly, in order to compensate for the spreading step, a diagonal correction is needed:
\begin{equation}
\widetilde B_\mathrm{pix}  = p_\mathrm{pix} b_\mathrm{pix}~,
\end{equation}
where the correction factors $p_\mathrm{pix}$ come from samples of the kernel Fourier transform. The grid size of the FFT coordinates $b_p$ scales according to the kernel and the input dimensions $b_n$.

The \texttt{cuFINUFFT} library~\citep{cufinufft} includes implementations for the type-1 and type-2 NUFFT but not for type-3. We use a modified version of \texttt{cuFINUFFT}, where we implement the missing transform type. 
The procedure for the full image synthesis is shown in Algorithm \ref{alg:nufft_synthesis}.

\begin{algorithm}[h!]
\caption{NUFFT Synthesis}\label{alg:nufft_synthesis}
\begin{algorithmic}[1]
\Procedure{collect}{$N_\mathrm{eig}, W, S, I, P_\mathrm{icrs}, P_\mathrm{uvw}, X_\mathrm{pix}$}
  \State $G \gets $ Compute gram matrix $(W, P_\mathrm{icrs})$
  \State $\alpha, \lambda \gets $ Generalized eigenvalue decomposition $(S, G)$
  \State $\alpha, \lambda \gets $ Select $N_\mathrm{eig}$ largest eigenvalues ($\alpha, \lambda$)
  \For{$n_f\in\{1,...,N_f\}$}
  \For{$n_i\in\{1,...,N_i\}$}
     \State $\alpha_I, \lambda_I \gets $ Eigenvalues / vectors in interval $I^{(n_i)}$
     \State $\lambda_f \gets$ Apply filter ($n_f, \lambda_I$)
     \State $V \gets  \alpha_I $ $\Call{diag}{\lambda_f} \alpha_I^H$ 
     \State $B^{(n_i, n_f)} \gets B^{(n_i, n_f)} + \Call{NUFFT}{V, P_\mathrm{uvw}, X_\mathrm{pix}}$
  \EndFor
  \EndFor
  \State Store $B$
\EndProcedure
\end{algorithmic}
\end{algorithm}

\subsubsection{Domain Partitioning}

\begin{figure*}
\centering
\begin{subfigure}{.5\linewidth}
  \centering
  \includegraphics[width=0.9\linewidth]{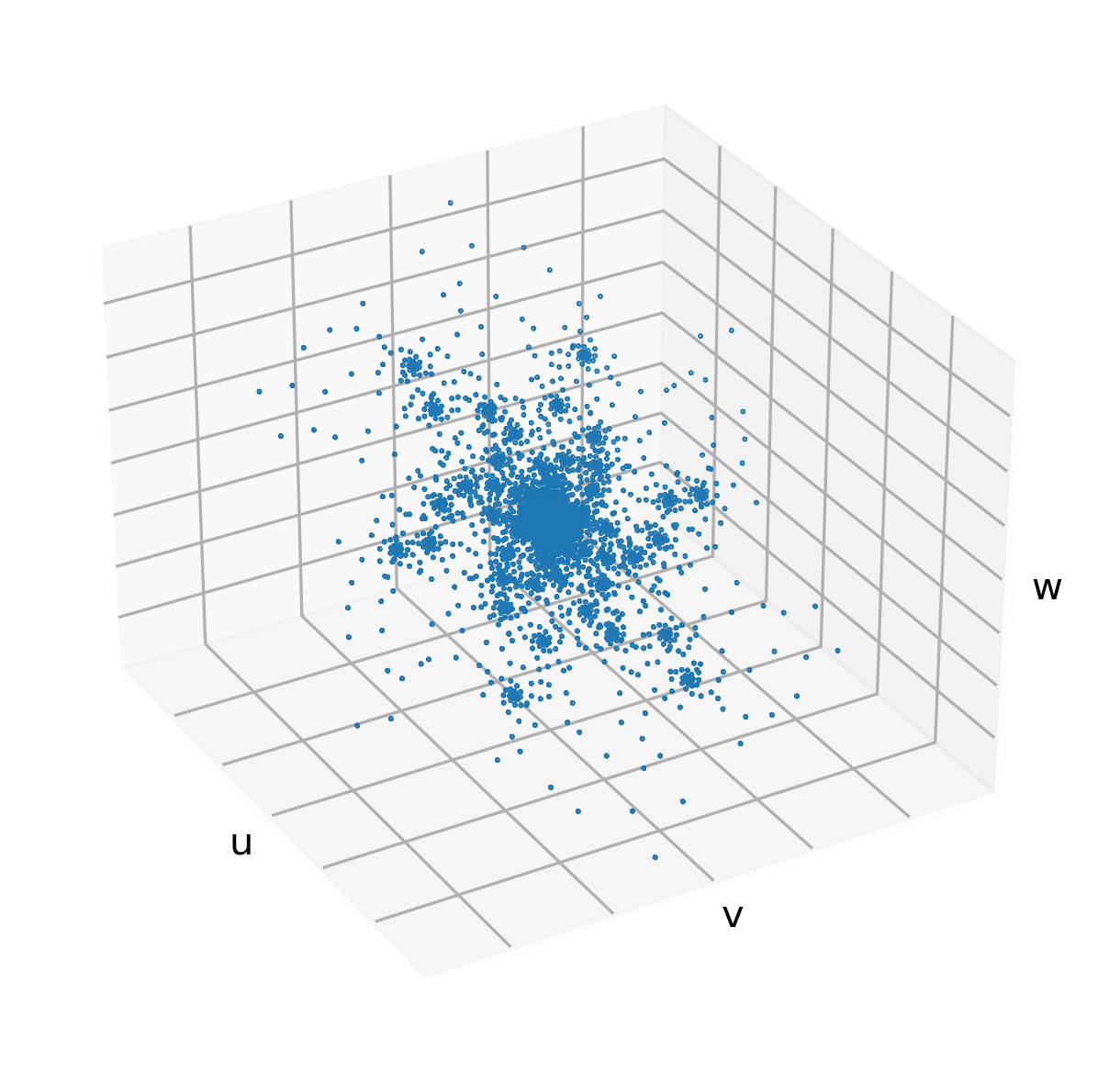}
  \caption{}
\end{subfigure}%
\begin{subfigure}{.5\linewidth}
  \centering
  \includegraphics[width=0.9\linewidth]{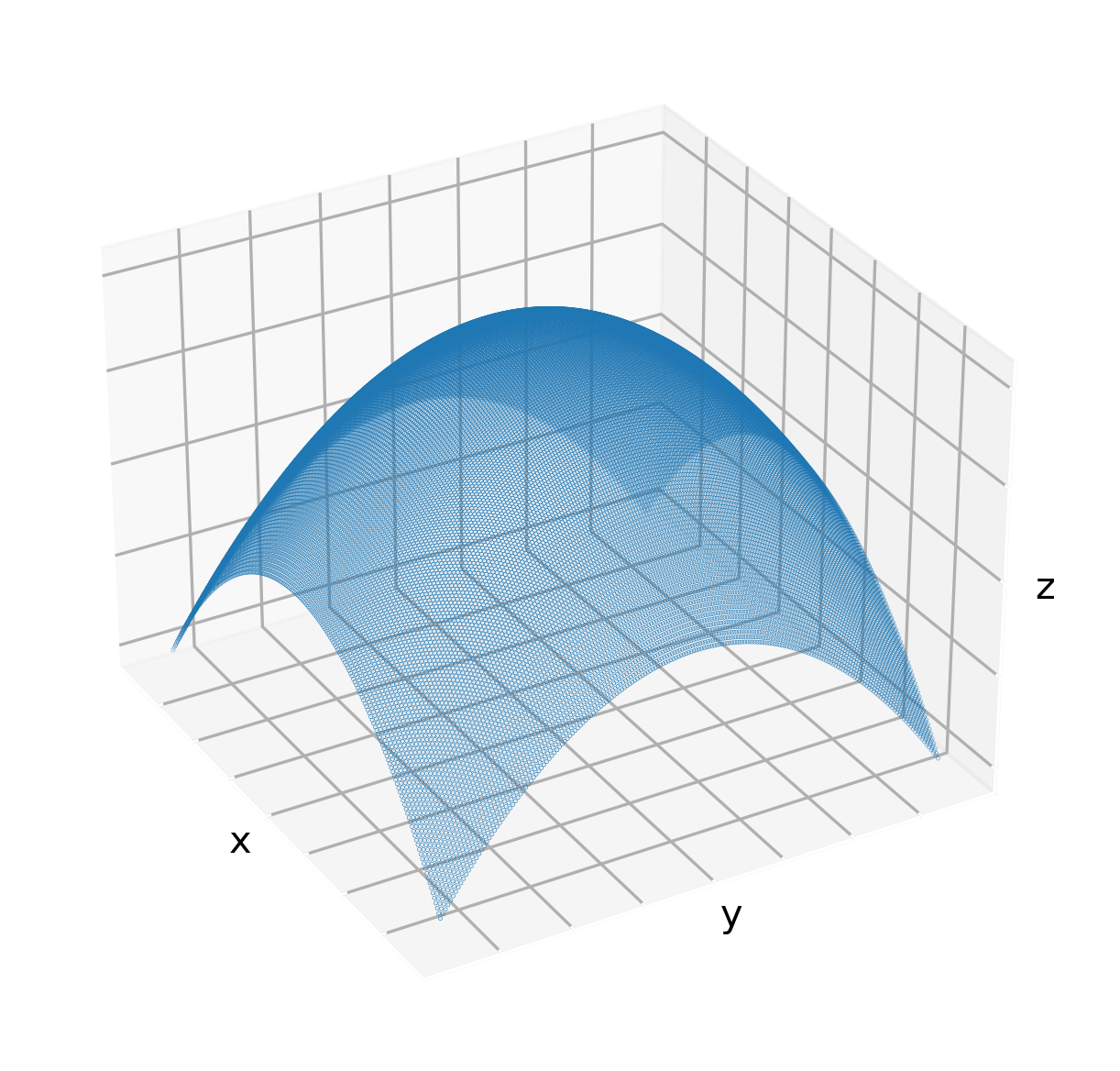}
  \caption{}
\end{subfigure}

\begin{subfigure}{.5\linewidth}
  \centering
  \includegraphics[width=\linewidth]{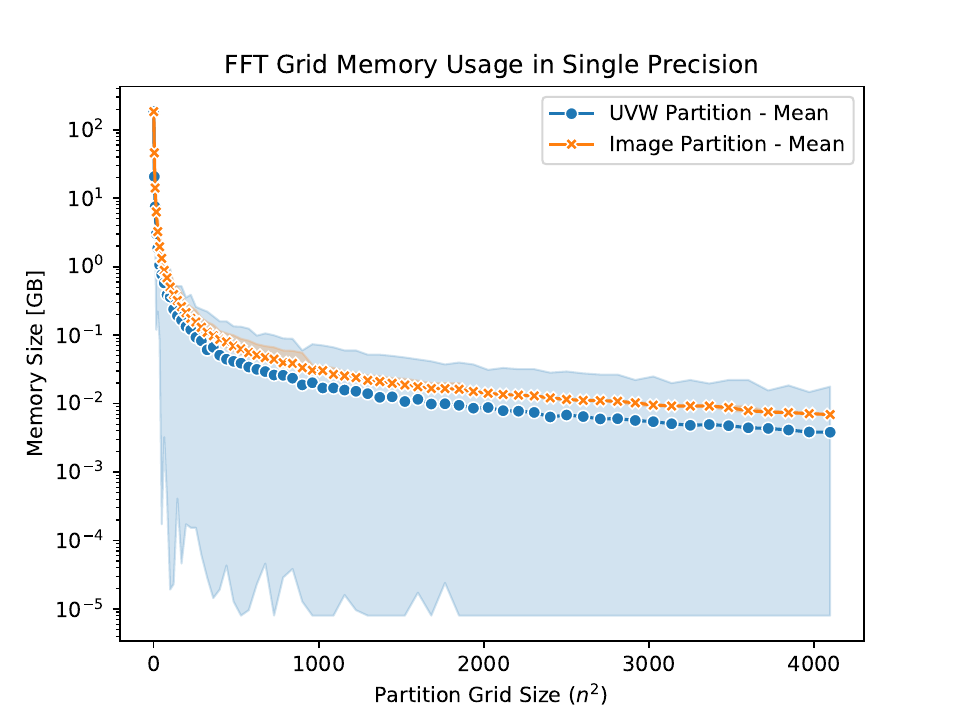}
  \caption{}
  \label{fig:ska_fft_memory}
\end{subfigure}%
\begin{subfigure}{.5\linewidth}
  \centering
  \includegraphics[width=\linewidth]{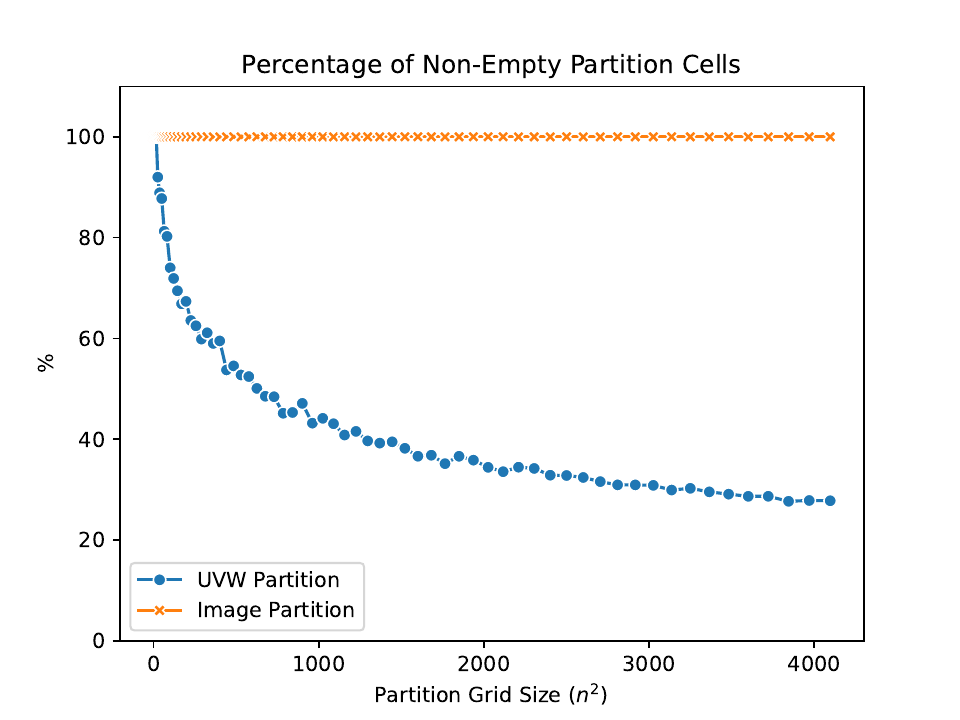}
  \caption{}
  \label{fig:ska_non_zero}
\end{subfigure}
\caption{Domain partitioning of the input uvw (a) and output image pixel (b) coordinates for a simulated SKA-Low antenna array. (c) shows the mean memory size of the uniform FFT grid used for computing the NUFFT for a domain partition grid of size $(n, n, 1)$. The coloured area indicates the total spread between the minimum and maximum size. (d) shows the fraction of non-empty cells in the partition grid.}
\label{fig:domain_part}
\end{figure*}

\label{subsec:partitioning}
Computing a NUFFT requires spreading the nonuniform points onto a uniform 3D grid, allowing us to calculate the FT using an implementation of the Fast Fourier Transform. The size of the grid is directly proportional to the extent of the input and output coordinates. Therefore, the efficiency of an NUFFT approximation compared to the direct evaluation of the sum in Equation~\ref{eq:type3nufft} depends on the spatial distribution of the input and output. 

Tightly clustered data is typically most suited for NUFFTs. However, {  data with multiple clusters of input coordinates}, such as certain antenna configurations used in radio astronomy, can be sub-optimal. In some cases, the required grid size may even exceed the available system memory.

 To optimize the NUFFT performance, we implement spatial domain partitioning of the input and output data which splits a single NUFFT into multiple independent NUFFTs, such that the FFT grid size of each individual transform can be kept as small as possible, a similar strategy leveraged by ~\cite{hvox2023}. For input partitioning into $I$  domain cells, this is equivalent to splitting the sum for each pixel into multiple sums:
\vspace{-0.5em}
\begin{equation}
\widetilde B_\mathrm{pix} = \sum_{n=1}^{N_1} V'_n e^{i x_\mathrm{pix}' \cdot b_n'} + ... + \sum_{n=N_{I-1}+1}^{N_I} V'_n e^{i x_\mathrm{pix}' \cdot b_n'}~.
\end{equation}
When combined with output partitioning into $O$ domain cells, the total number of NUFFTs is $O\times I$, where the pair-wise combination of input and output domain cells provides the input and output of a NUFFT.

We implemented domain partitioning using a regular 3D grid for both input and output. Figure \ref{fig:domain_part} shows the results for simulated SKA-Low data with partitioning of the input UVW domain and output image domain. The input data consists of clustered points with some individually scattered around the domain, while output points are evenly spaced on a curved surface.

In Figure \ref{fig:ska_fft_memory}, the memory usage per FFT is shown using a partitioning grid of size $(n, n, 1)$. For output partitioning of the more evenly spaced points, there is barely any difference in FFT grid size between each domain cell. In contrast, there is a wide variability in FFT grid size for the input UVW partitioning. This is also reflected in Figure~\ref{fig:ska_non_zero}, which shows the percentage of domain cells containing any data points. For large partition grid sizes, as little as 30\% of the domain cells require the computation of a NUFFT, which may, therefore, reduce the total number of operations required. However, a NUFFT will also require the spreading of input data and interpolation to output data, which only partially benefit from a smaller FFT grid size.

Whether domain partitioning provides a performance benefit is, therefore, highly dependent on the input and output distribution. It does, however, provide an avenue for parallelization and, as shown in Figure~\ref{fig:ska_fft_memory}, can reduce memory usage from over 200 GB to less than 1 GB.

\subsection{Image weighting}
\BIPP can support ``natural'' and ``uniform'' image weighting by applying appropriate weights to the input visibilities before eigenvalue decomposition. For ``natural'' weighting, no additional weights are applied to the image. The system noise is smallest in this case, but resolution and sidelobe noise are worse. For ``uniform'' weighting, we calculate the sampling density function on a grid in $uv$ space with $\Delta u = \Delta v = 1/\text{FoV}$. Each visibility is then weighted with the inverse of its sampling density function. Unless otherwise noted, the results in Sections~\ref{sec:val}, ~\ref{sec:perf}, and ~\ref{sec:SA} are created using natural image weighting.

\subsection{Current Limitations}
\label{sec:limitations}

{  As discussed in Section~\ref{sec:vsclean}, the images produced by the }Bluebild algorithm do not include any  regularization. As such, the images produced by \BIPP are comparable to the  ``dirty'' images produced by standard radio astronomy imaging algorithms, i. e. the true image of the sky is convolved with the instrument's point spread function~\citep{synthesisimaging1999}. Additional deconvolution { would be} required to produce a comparable ``clean'' image. Nevertheless, the eigenvalue decomposition offers some interesting advantages, as discussed in Section~\ref{sec:SA}.

\BIPP assumes that the array contains calibrated omnidirectional antennas $g_p(\vec r) = 1$.  Future developments will include an option to include a non-uniform instrument response.

{ Currently \texttt{BIPP} uses all baselines to reconstruct the image}, even if those baselines provide a resolution higher than the imaging resolution.  In other radio astronomical imaging libraries, these longer baselines are discarded to improve performance. The effect of baseline truncation must be kept in mind when comparing \texttt{BIPP} to other imaging software. { An option for baseline truncation will be added in a future release}.

Finally, while \texttt{BIPP}  has been extensively validated against data files from MWA and LOFAR, as discussed in Section~\ref{sec:val}, we have not exhaustively tested input files from every telescope. {  To include beamforming as described in Section~\ref{sec:beamforming}, \texttt{BIPP} reads in time-varying  station element positions}, which are not necessarily structured the same in \texttt{CASA} MeasurementSet files from different telescopes.

\section{Validation}
\label{sec:val}
To validate our implementation of Bluebild, we compare the outputs of \bipp with the outputs of the original Python implementation. Furthermore, we compare the consistency of our implementations of Standard Synthesis with NUFFT synthesis and the CPU and GPU implementations. In total, we compare five different implementations of the Bluebild algorithm:
\begin{itemize}
\item \texttt{BluebildSsCpu}: Original implementation in Python of the Standard Synthesis algorithm from the \texttt{pypeline} library~\citep{KashaniBluebild}, running on CPU
\item \verb|BippSsCpu|: C++ implementation of the Standard Synthesis algorithm, running on CPU
\item \verb|BippSsGpu|: C++ implementation of the Standard Synthesis algorithm, running on GPU
\item \verb|BippNufftCpu|: C++ implementation of the NUFFT Synthesis algorithm, running on CPU
\item \verb|BippNufftGpu|: C++ implementation of the NUFFT Synthesis algorithm, running on GPU
\end{itemize}
For the validation tests, the NUFFT user-requested tolerance $\epsilon$ is set to $10^{-5}$.
\label{sec:valintro}

Furthermore, we compare the \BIPP output images to those produced with two reference packages in the field: \texttt{WSClean} (version 3.4, using the \texttt{wgridder} gridder with its default accuracy of 1e-4) \citep{offringa-wsclean-2014} and the \texttt{tclean} task from \texttt{CASA} (version 6.6.3-22, using the \texttt{wproject} gridder, setting the number of distinct w-values to -1, letting this parameter to be automatically defined by \texttt{CASA}) \citep{casa2007}. For both of these imaging libraries, the number of cleaning iterations is set to zero. Thus, we only compare dirty images {  $\hat I$} to \BIPP outputs {  $\widetilde I$}.

\subsection{Dataset}
\label{sec:valdata}
For our validation checks, we use \texttt{OSKAR} (\cite{OSKAR}, release 2.8.3) to create four simulated SKA-Low telescope observations with a configuration of 512 stations and variable field-of-view (FoV) of  17, 34, 68 or 136 arcmin. We create a radio sky  with 9 point sources of 1 Jy spread over a regular grid\footnote{Sources were simulated to be located at the center of the pixels located at 1/8, 4/8 and 7/8 of the image width on both images axes.} over the FoV.
50 time steps are generated spread over an observing period of 6 hours.

The simulated visibility data are then processed with \texttt{CASA}, \texttt{WSClean} and all five implementations of Bluebild. In all cases, no cleaning is performed, and Bluebild LSQ images $\widetilde{I}$ are compared to the \texttt{CASA} and \texttt{WSClean} so-called "dirty" images. To ensure a fair comparison, the resolution is fixed to 4 arcsec, so to prevent the visibility truncation of \texttt{WSClean} and \texttt{CASA} as discussed in Section~\ref{sec:limitations}. This results in square images with $256^2$, $512^2$, $1024^2$ and $2048^2$ pixels. 

These same datasets and imaging strategies are used for the performance benchmarks described in Section~\ref{sec:perf}.

\subsection{Impact of the energy clustering}
\label{val:subsec:impact_energy_clustering}
For each of the five implementations of Bluebild 
we assess the impact of the { eigenvalue partitioning} (see Sect.~\ref{subsec:energy_levels_and_partitioning}) on image reconstruction. All eigenvalues (positive and negative) are considered, with no eigenvalue truncations, such that $N_\text{eig} = N_A$. Eigenvectors $\{\alpha_a \}$ with positive eigenvalues are clustered into 1, 2, 4, or 8 distinct energy levels, and all eigenvectors with negative eigenvalues are { partitioned} into a single separate layer. { The resulting energy levels are summed} together to create the complete LSQ image $\widetilde{I}$.

\begin{table}\centering
\footnotesize
\begin{tabular}{@{}lrccc@{}}\toprule
Solution      &   Im. width  &   Levels   &   Maximal range  & RMS       \\
              &   [pixel]     &            &   [Jy/beam]  & [Jy/beam] \\ \midrule
\texttt{BluebildSsCpu} &  256 & 1 \& 4 & [-4.8e-7, 4.8e-7] & 8.3e-8 \\
\texttt{BippSsGpu    } & 2048 & 1 \& 2 & [-2.0e-6, 7.2e-7] & 2.4e-8 \\
\texttt{BippSsCpu    } & 2048 & 1 \& 2 & [-2.1e-6, 7.7e-7] & 2.4e-8 \\
\texttt{BippNufftGpu } &  256 & 1 \& 8 & [-3.4e-5, 2.6e-5] & 1.0e-5 \\
\texttt{BippNufftCpu } &  256 & 1 \& 8 & [-1.3e-6, 1.5e-6] & 2.7e-7 \\
\bottomrule
\end{tabular}
\caption{Maximal range of intensity differences reported for each implementation of Bluebild when comparing images obtained using different number of energy levels for the clustering of the eigenvalues.}
\label{val:tab:impact_of_energy_clustering}
\end{table}

\begin{figure}
    \centering
    \includegraphics[width=0.4\textwidth]{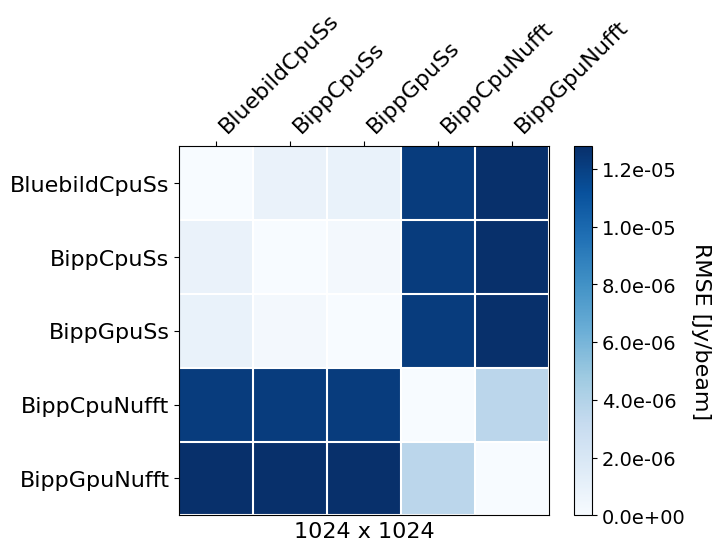}
    \caption{Inter-solution consistency between the different implementations of Bluebild for images of $1024 \times 1024$ pixels and produced by combining all positive eigenvalues into a single layer. Similar results are obtained when clustering eigenimages into 2, 4 or 8 positive energy levels, see Section~\ref{subsec:energy_levels_and_partitioning}. 
    }.
    \label{fig:val:consistency_nlev_1_size_1024}
\end{figure}

For each implementation and each image size we assess the impact of the energy clustering by computing differences between images. Results are summarized in Table~\ref{val:tab:impact_of_energy_clustering}, which gathers, for each solution, the largest interval of differences between two energy levels over all image sizes. %
We emphasize here that Table~\ref{val:tab:impact_of_energy_clustering} reports the worst-case scenarios for each implementation. We find that the energy clustering scheme has a negligible impact on the resulting image, as expected from Eq.~\ref{eq:eigenvaluedeco}. The errors observed are consistent with single-point floating precision used in the calculation for Standard Synthesis implementations and consistent with the tolerance $\epsilon = 10^{-5}$ tolerance for NUFFT implementations, as expected. %

\subsection{Consistency between implementations}
\label{val:subsec:consistency_between_implementations}

\begin{figure*}
     \centering
     \begin{subfigure}[b]{0.25\textwidth}
         \centering
         \includegraphics[width=\textwidth]{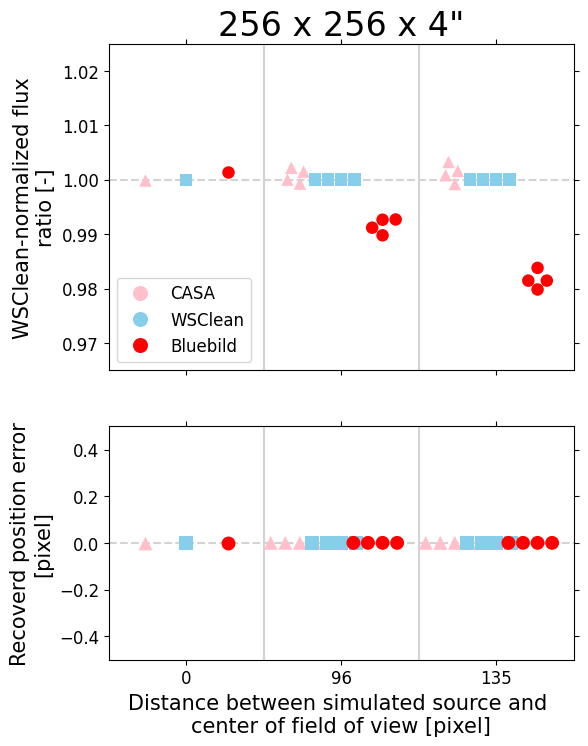}
     \end{subfigure}
     \hfill
     \centering
     \begin{subfigure}[b]{0.23\textwidth}
         \centering
         \includegraphics[width=\textwidth]{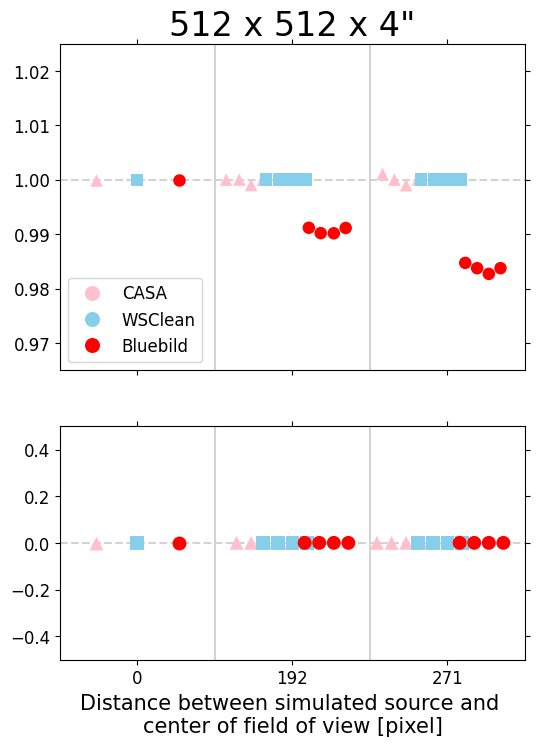}
     \end{subfigure}
     \hfill
     \begin{subfigure}[b]{0.23\textwidth}
         \centering
         \includegraphics[width=\textwidth]{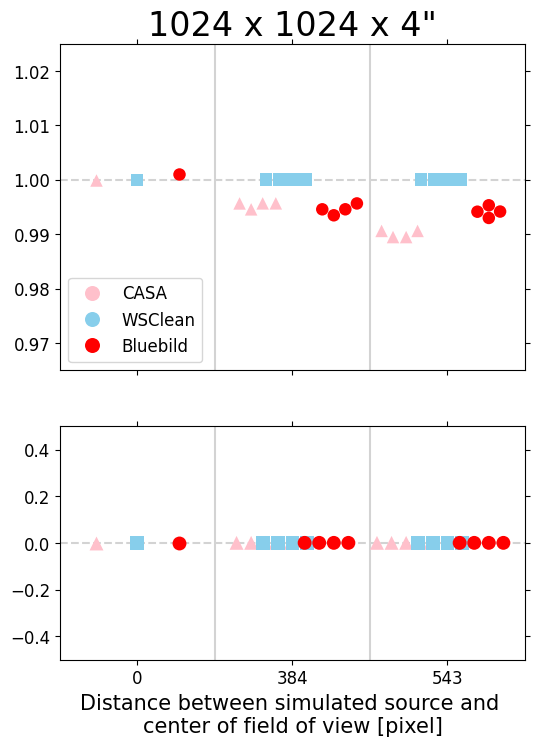}
     \end{subfigure}
     \hfill
     \begin{subfigure}[b]{0.23\textwidth}
         \centering
         \includegraphics[width=\textwidth]{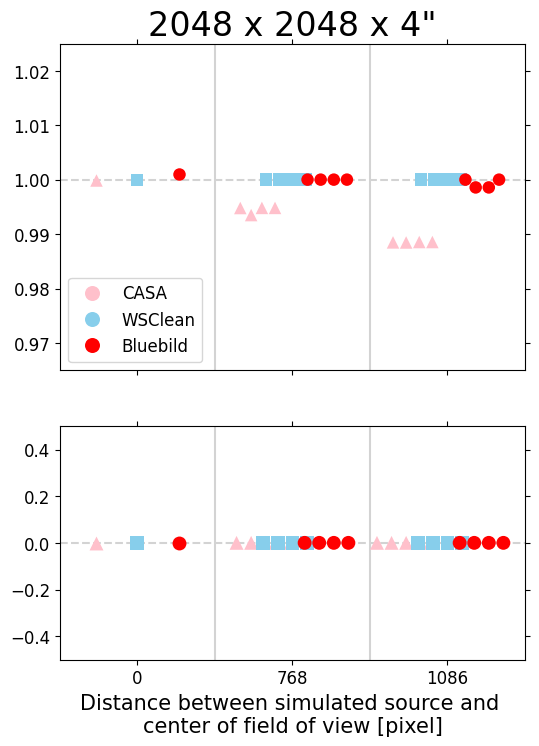}
     \end{subfigure}
        \caption{Top plots: { recovered intensities relative to the flux measured in the \texttt{WSClean} image and bottom plots: distance between the recovered source positions and the simulated ones, in pixels, all plotted against the distance in pixels to the centre of the field of view. The size of the images increases from the left to the right from 256 to 2048 pixels.  The Bluebild solution (in red) was obtained with \texttt{BIPP} NUFFT GPU implementation. Inconsistency in recovered flux as a function of image size is likely due to gridding/interpolation effects because we use the maximum pixel value for the recovered intensity.}}
        \label{fig:validation_new:recovered_sky:point_sources_intensities_distances}
\end{figure*}

We evaluate the consistency of the images produced by our five implementations of Bluebild. Consistency here is measured as the RMS of pixel intensity differences between two images, referred to as the RMS error or RMSE. Figure~\ref{fig:val:consistency_nlev_1_size_1024} presents the inter-solution consistency for the $1024^2$ pixel image. 

The overall agreement between all solutions is excellent, with a quasi perfect agreement between the three  Standard Synthesis solutions while the maximum RMSE of only 1.28e-05 Jy/beam is found between the \texttt{BippNufftGpu} and Standard Synthesis. Agreement between \texttt{BippNufftCpu} and the Standard Synthesis solutions reduces to 1.22e-05 Jy/beam while the agreement between \texttt{BippNufftGpu} and \texttt{BippNufftCpu} is around 3.61e-06. These numbers were obtained for $1024^2$ pixel images.

\subsection{Recovering simulated point sources}

\begin{figure*}
    \centering
         \includegraphics[width=0.92\textwidth]{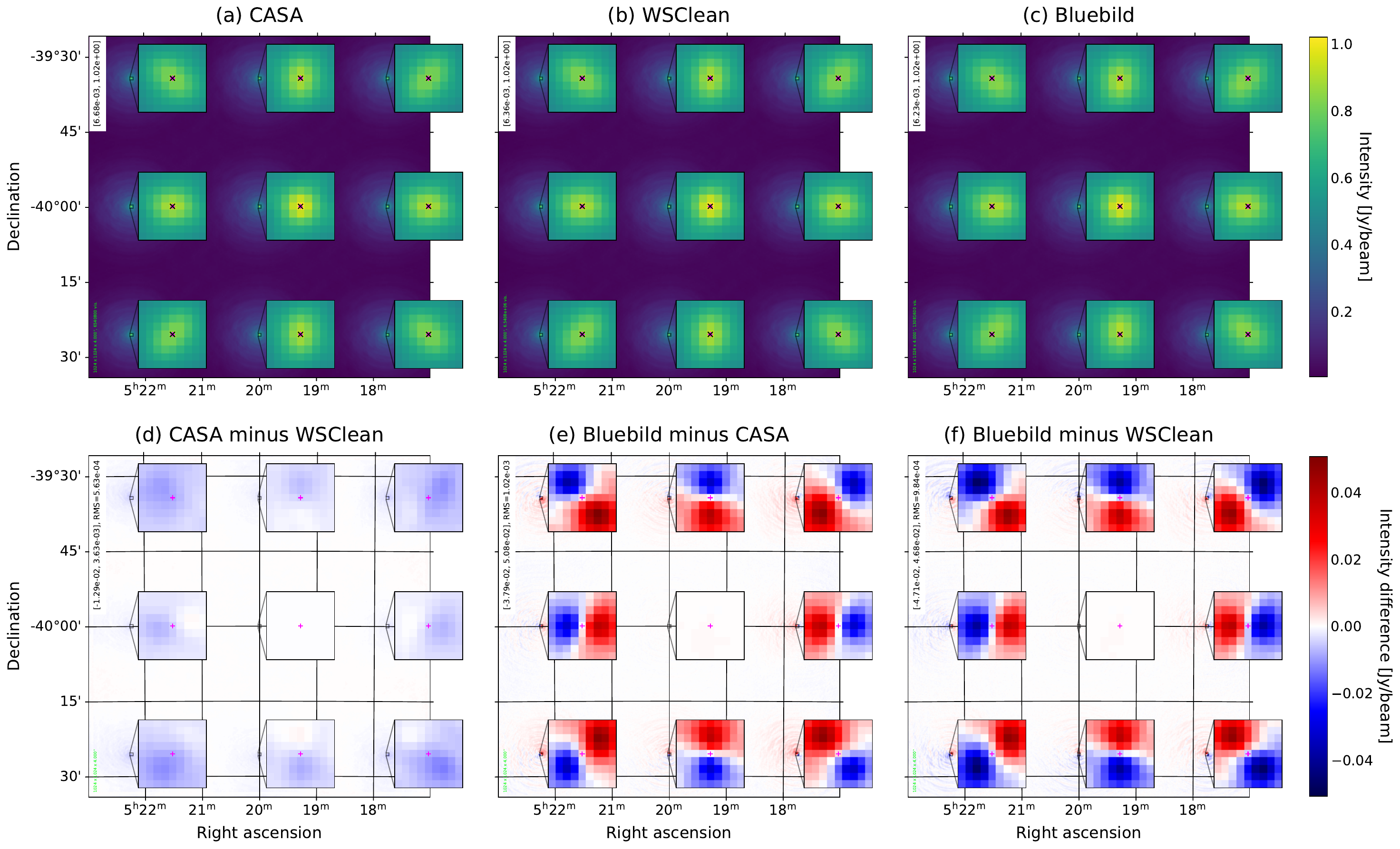}

    
    \caption{Recovered simulated 1 Jy point sources by \texttt{CASA}, \texttt{WSClean} and \texttt{BIPP} and differences between output images. Images are of $1024 \times 1024 \times 4$" resolution. Pink crosses indicate the positions of the simulated sources whereas the black ones indicate the position of recovered sources, i.e. the position of pixel of highest intensity in the vicinity of the true position. The \texttt{BIPP} solution was obtained with the NUFFT Synthesis algorithm and running on GPU.}
    \label{fig:validation_point_sources_maps}
\end{figure*}

\begin{figure*}
    \centering
    \includegraphics[width=0.92\textwidth]{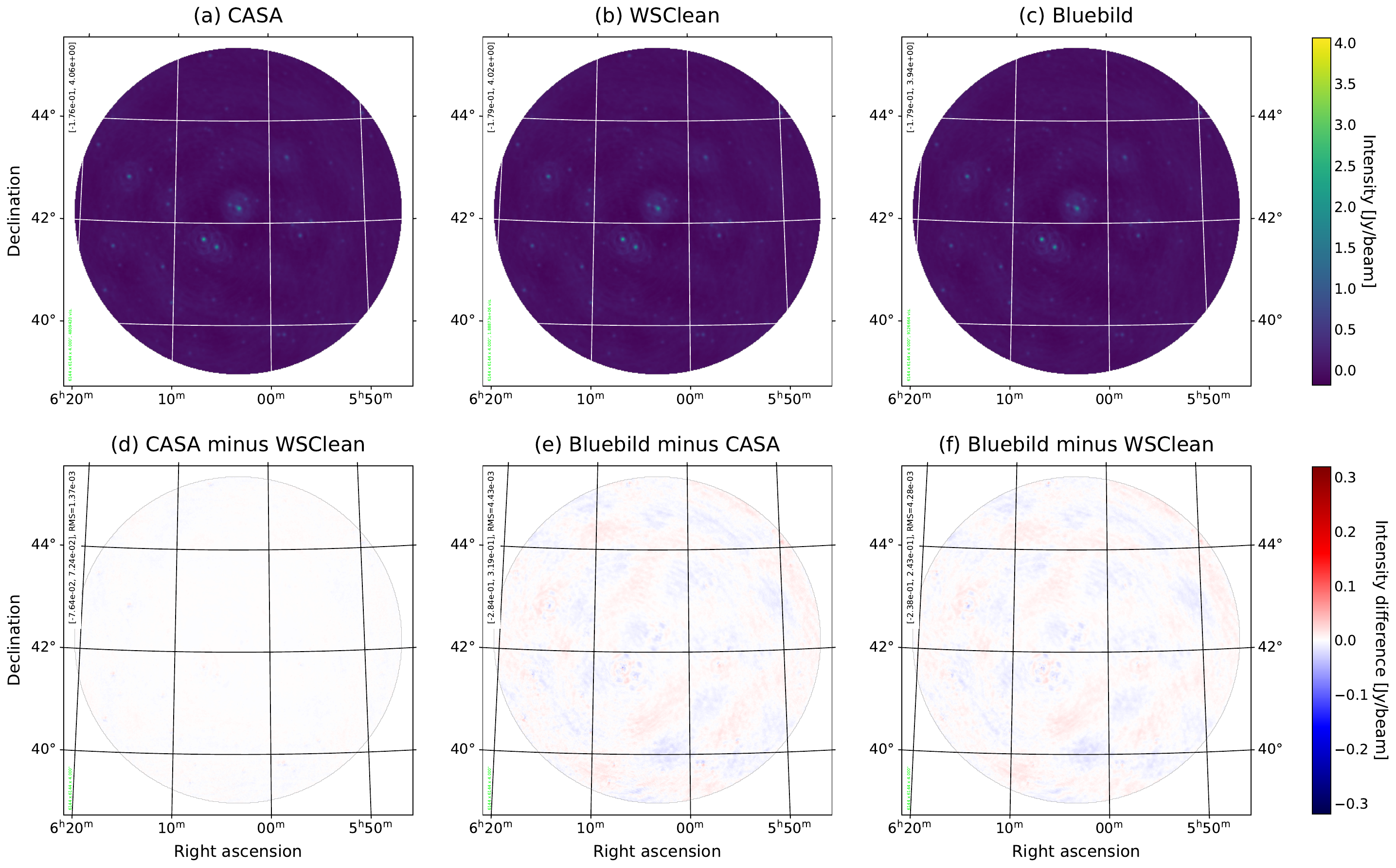}
    \caption{Top row: dirty maps produced with (a) \texttt{CASA}, (b) \texttt{WSClean}, (c) \BIPP using real LOFAR data from the Toothbrush cluster RX J0603.3+4214 dataset from \protect\cite{LEAP}. Images produced are 6144 by 6144 pixels of $4"$ angular resolution. Bottom row: differences between the following pairs of dirty maps: (d) \texttt{CASA} minus \texttt{WSClean}, (e) \bipp minus \texttt{CASA} and (f) \bipp minus \texttt{WSClean}. Mask from the \texttt{CASA} solution was applied to the two others for a fair comparison.}
    \label{fig:val_new:real_data_lofar}
\end{figure*}

In addition to checking the per-pixel consistency between imaging solutions, we also evaluate the ability of the three packages \texttt{CASA}, \texttt{WSClean}, and \texttt{BIPP} to recover properties of the simulated point sources in our validation dataset. We calculate
the distance (in pixels) between recovered and simulated sources' positions and the intensity of the recovered sources compared to the simulated ones (1 Jy everywhere). 
In Figure~\ref{fig:validation_point_sources_maps} we show the outputs of \texttt{CASA}, \texttt{WSClean}, and \texttt{BIPP} imaging the same visibilities to create $1024^2$ pixel images with 4" resolution. Pink crosses indicate the true/simulated positions of the point sources, and  dark crosses indicate their recovered positions, taken as the position of the pixel of highest intensity in the vicinity of the location of the simulated source ($9 \times 9$ pixels square area centered on the source's true location). 

Figure~\ref{fig:validation_new:recovered_sky:point_sources_intensities_distances} provides a global summary of the recovery of the point sources for the \texttt{BippNufftGpu} solution for all image sizes. { All three software packages recover consistent source positions and { intensities. In the} field comparisons in Figure~\ref{fig:validation_point_sources_maps} we see that the \texttt{BIPP} image has almost perfect agreement in the center of the FoV, but sources farther from the center are slightly offset with respect to the images reconstructed by \texttt{CASA} and \texttt{WSClean}, with offset increasing with distance. This is likely due to the use of the 3D NUFFT for imaging directly on the sphere instead of a $w$-term approximation. These sub-pixel offsets do not change the recovered source position.}

\subsection{Processing real LOFAR data}
Finally, we also check the output of \BIPP on real data collected by the Low-Frequency Array (LOFAR) telescope.
We use dataset IV from \cite{LEAP}, a LOFAR observation of the “Toothbrush” cluster RX J0603.3+4214 from 36 stations over the period 2013-02-24-15:32:01.42 to 2013-02-25-00:09:24.51, resulting into 3,123 observation { time steps. We use just a single frequency channel with central frequency of 130.2 MHz}. Dirty images of $6144 \times 6144$ pixels of 2" angular resolution were computed with \texttt{CASA} and \texttt{WSClean}, and compared to the LSQ image $\widetilde{I}$ produced by \texttt{BIPP}. The \texttt{BIPP} solution was obtained with the \texttt{BippNufftGpu} implementation with a convergence criterion of $\epsilon = 10^{-5}$.

The results are shown in Figure~\ref{fig:val_new:real_data_lofar}.
The overall similarity between the three solutions is excellent, with maximal differences at the 2.5\% level found in the comparison between the \texttt{CASA} and \texttt{BIPP} located around the brightest source.  \texttt{CASA} and \texttt{WSClean} images show the best agreement, with an RMSE of 1.65e-3 Jy/beam. Differences between \texttt{BIPP} and the two other solutions are a bit more pronounced { compared to the simulated dataset}, with large scale light structures, resulting is larger RMSE of 4.92e-3 (\texttt{CASA}) and 4.70e-3 (\texttt{WSClean}). { This is likely because this field is less sparse than the simulated dataset with only 9 sources.}

{ 
\subsection{LSQ Imaging}
The differences between the \texttt{BIPP} and \texttt{CASA}/\texttt{WSClean} images is mostly driven by the different gridding algorithm that these software packages use for imaging. \texttt{WSClean} uses w-stacking~\citep{offringa-wsclean-2014}, \texttt{CASA} uses w-projection~\citep{wproj2008}, and \texttt{BIPP} uses a 3D NUFFT of type 3 (non-uniform input to non-uniform output).

To disentangle these effects, we also evaluate the difference between the LSQ-consistent image $\widetilde I$ constructed by Bluebild with the back-projected dirty image $\hat I$, using the 3D NUFFT of type 3 to perform the gridding and FFTs in both cases. The results for both the simulated data with 9 point sources and the LOFAR Toothbrush data are shown in Figure~\ref{fig:validation_new:gramcheck}. As discussed in Section~\ref{sec:vsclean}, the Gram matrix has a small effect if the instrument baselines are much larger than the observing wavelength, which is true for both of these datasets. The differences we see between $\widetilde I$ and $\hat I$ are at the sub 0.1\% level, which is smaller than the differences observed between \texttt{CASA} and  \texttt{WSCLEAN} reconstructed images. The \texttt{CASA} - \texttt{WSCLEAN} RMS is $5.6\times10^{-4}$ Jy/beam  for the simulated SKA-Low data and $1.4\times10^{-3}$ Jy/beam for the LOFAR data, which is larger than the RMS for the $\widetilde I - \hat I$ residual of $1.1\times10^{-4}$ Jy/beam and $8.29\times10^{-5}$ Jy/beam, respectively.

\begin{figure*}
     \centering
     \begin{subfigure}[b]{0.45\textwidth}
         \centering
         \includegraphics[width=\textwidth]{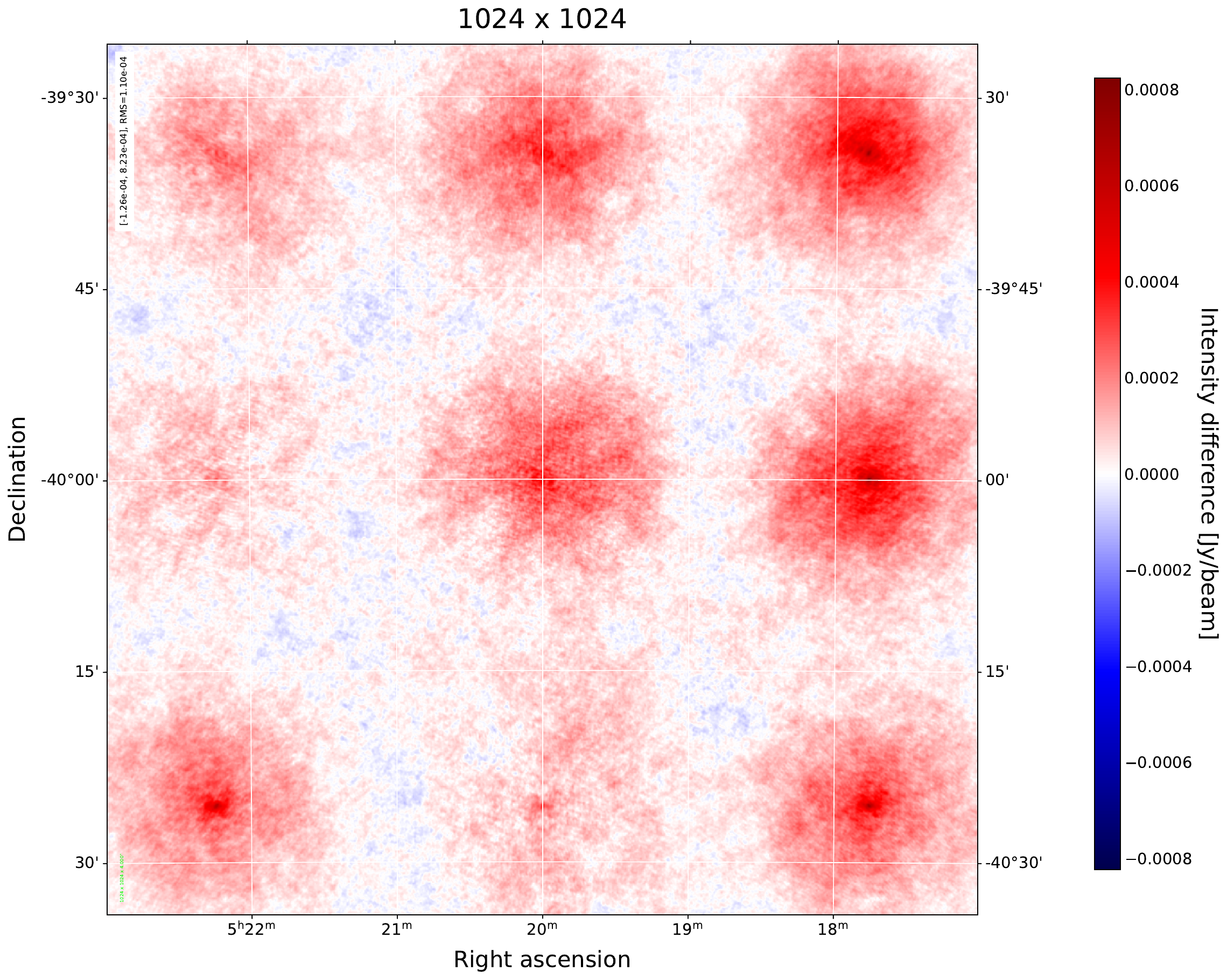}
     \end{subfigure}
     \hfill
     \centering
     \begin{subfigure}[b]{0.45\textwidth}
         \centering
         \includegraphics[width=\textwidth]{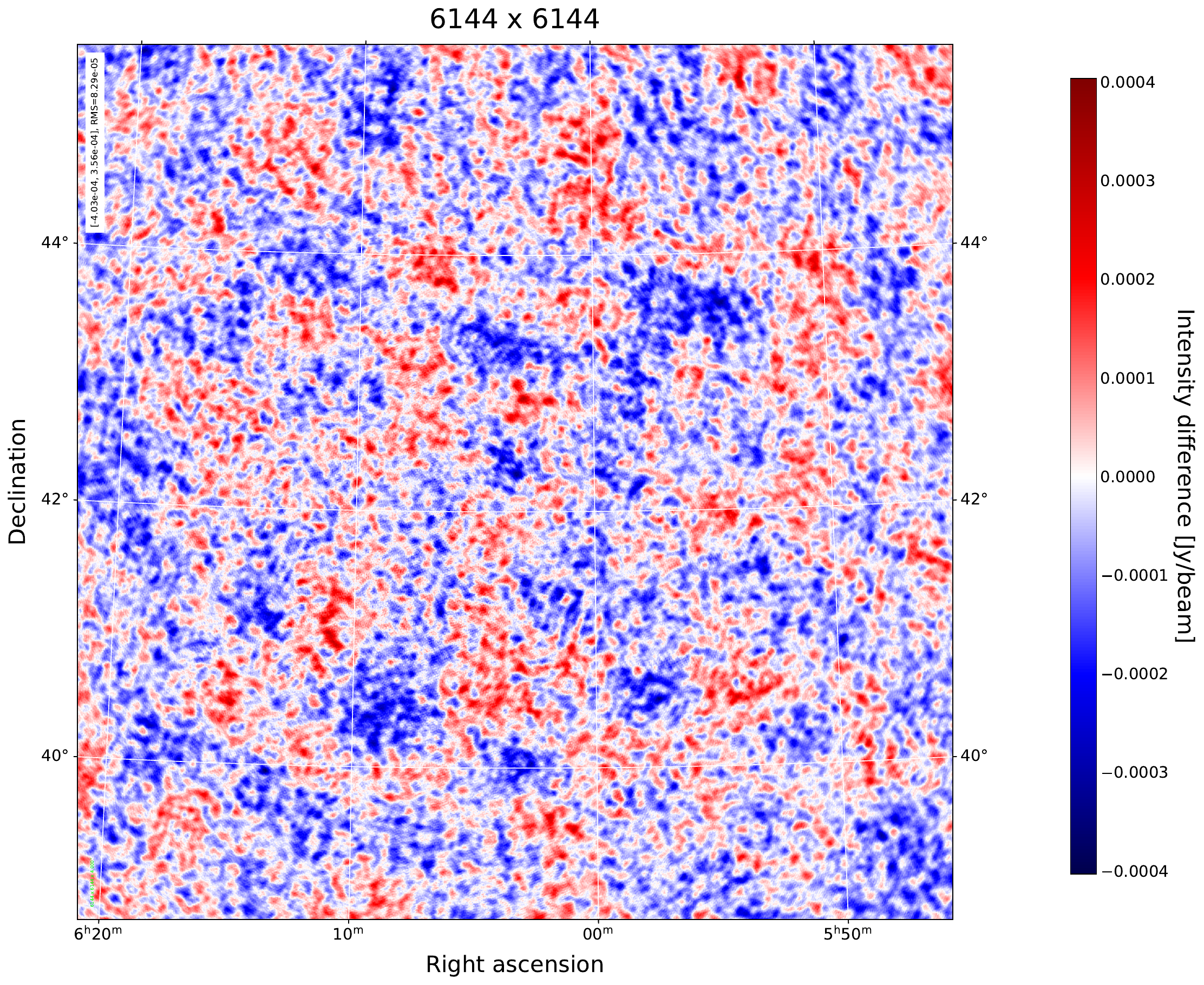}
     \end{subfigure}
        \caption{ $\hat I - \widetilde I$ for (left) simulated SKA-Low observation with 9 sources and (right) the Toothbrush cluster RX J0603.3+4214 LOFAR observation. Observed differences are at the sub 0.1\% level, smaller than the differences observed between \texttt{CASA} and  \texttt{WSCLEAN}. }
        \label{fig:validation_new:gramcheck}
\end{figure*}
}

\section{Performance Tests}
\label{sec:perf}
In this section we examine the computational performance of the five implementations of the Bluebild algorithm (i.e. \texttt{BluebildSsCpu}, \texttt{BippSsCpu}, \texttt{BippSsGpu}, \texttt{BippNufftCpu} and \texttt{BippNufftCpu} as described in Section~\ref{sec:valintro}) to the reference packages \texttt{CASA} and \texttt{WSClean}. As in Section~\ref{sec:val}, we only compare the execution times for \texttt{CASA} and \texttt{WSClean} to create dirty images.

For the benchmark, we use the same sets of data as described in Section~\ref{sec:valdata}. We process 50 { time steps} of \texttt{OSKAR} simulated SKA-Low observations based on 512 observing stations. The pixel angular resolution was fixed to 4 arcsec while the size of the image was set to either $256^2$, $512^2$, $1024^2$ or $2048^2$ pixels, resulting in a varying FoV. For each of these image sizes, all five Bluebild implementations were run with 1, 2, 4 and 8 positive energy levels on top of a single negative energy level containing all of the negative eigenvalues. So, in total, our benchmark contains, for each image size, \(4 \times 5 = 20\) Bluebild results.

All the computations were carried out using single-precision (32-bits) floating-point numbers. NUFFT partitioning for both uvw and image domains was set to (4, 4, 1), as described in Section~\ref{subsec:partitioning}. The code was compiled using the GCC v11.3.0\footnote{\href{https://gcc.gnu.org/gcc-11/}{https://gcc.gnu.org/gcc-11/}} and NVCC v11.8.89 \footnote{\href{https://docs.nvidia.com/cuda/cuda-compiler-driver-nvcc/index.html}{https://docs.nvidia.com/cuda/cuda-compiler-driver-nvcc/index.html}} compilers.

Each solution is run with exclusive access to one of the large memory nodes of the GPU cluster of EPFL called Izar. Each node has 384 GB of DDR4 RAM, 2 Intel Xeon Gold 6230 CPUs running at 2.10 GHz with 20 physical cores each (one thread per physical core), and 2 NVidia V100 PCIe 32 GB GPUs. Multi-threaded regions of the code use the 40 CPU cores available. GPU-accelerated parts of the code use a single GPU out of the two available.

\begin{figure*}
\centering
\begin{subfigure}{0.48\textwidth}
    \includegraphics[width=0.99\textwidth]{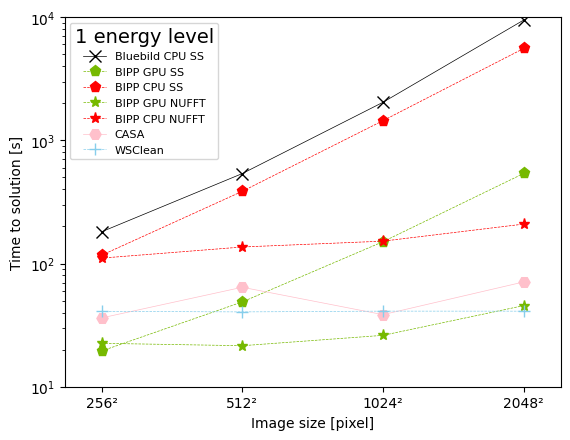}
\end{subfigure}
\hfill
\begin{subfigure}{0.48\textwidth}
    \includegraphics[width=0.99\textwidth]{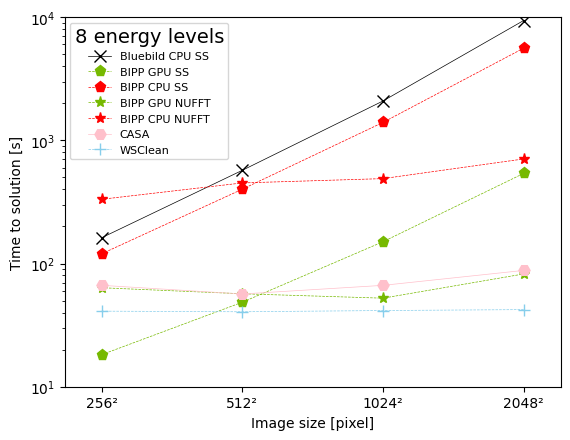}
\end{subfigure}
\begin{subfigure}{0.48\textwidth}
    \includegraphics[width=0.99\textwidth]{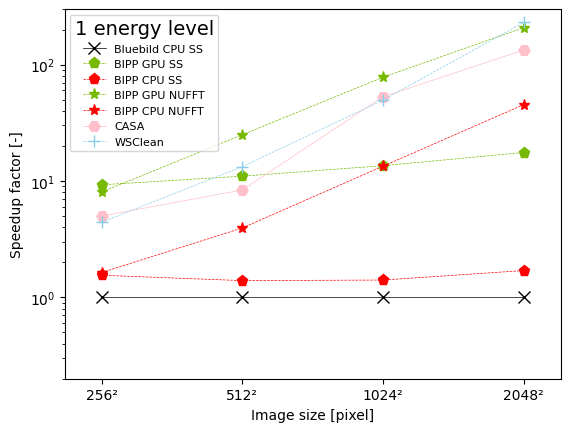}
\end{subfigure}
\hfill
\begin{subfigure}{0.48\textwidth}
    \includegraphics[width=0.99\textwidth]{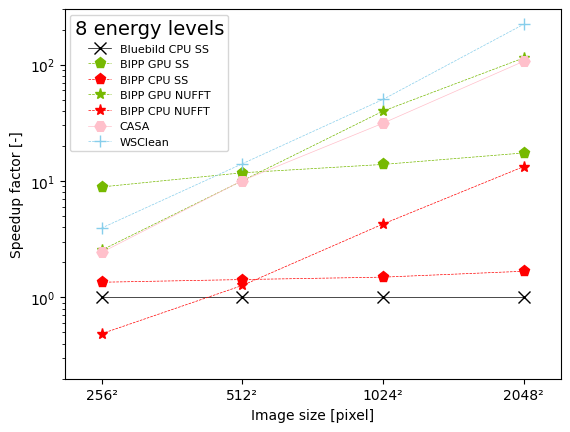}
\hfill
\end{subfigure}
\caption{Top row: times to solutions { when imaging with one or eight distinct energy levels. Bottom row: the} corresponding speedup factors computed over the reference \texttt{BluebildSsCpu} solution.}
\label{fig:perf:times_to_solutions__speedup_factors}
\end{figure*}

We analyse the times to solutions (TTS) and speedup factors (SF) obtained with the different \texttt{BIPP} implementations when compared to the reference Python implementation of Bluebild. Results are displayed in Figure~\ref{fig:perf:times_to_solutions__speedup_factors} for 1 energy level (left) and 8 energy levels (right).{  Table~\ref{tab:val:dirty_maps} provides the speedup factors of all \texttt{BIPP} solutions, \texttt{CASA} and \texttt{WSClean} when compared to \texttt{Bluebild} in producing images.  Table~\ref{tab:perf:tts} presents the decomposition of the time to solution for the best performing \texttt{BIPP} solution, namely the \texttt{BIPP} NUFFT GPU implementation when constructing the LSQ image using only a single  energy level. We notice the following results:
\begin{itemize}
\item  NUFFT Synthesis has better scaling than Standard Synthesis, as expected from the complexity analysis in Section~\ref{sec:complexity}.
\item Combining eigenvectors into distinct energy levels as discussed in Section~\ref{subsec:energy_levels_and_partitioning} improves the performance of NUFFT Synthesis by reducing the number of calls to the NUFFT, but does not impact the performance of Standard Synthesis.
\item BIPP GPU implementations outperform CPU implementations for both Standard Synthesis and NUFFT Synthesis.
\item\texttt{BippNufftGpu} is the fastest imager for images with $N_{pixel} \leq 1024^2$, faster than both   \texttt{WSClean} and \texttt{CASA}. However, the scaling appears to be better for \texttt{WSClean} and \texttt{CASA} and they perform better for larger image sizes.
\item The eigenvalue decomposition takes up a very small part of the total execution time, as shown in Table~\ref{tab:perf:tts}.
\end{itemize}
Most of the execution time of the NUFFT GPU implementation is dominated by the 3D NUFFT. However, it would be possible to use $w$-projection or $w$-stacking to evaluate the DFT of the Gram-corrected visibilities in Equation~\ref{eq:type3nufft}, allowing more performant fPCA-based LSQ imaging for larger image sizes.}
\begin{table}
\centering
\footnotesize
\setlength{\tabcolsep}{3pt} 
\renewcommand{\arraystretch}{1.0} 
\begin{tabular}{@{}lrrrrrrrrr@{}}\toprule
Solution        &  \multicolumn{9}{c}{Speedup factors [-]} \\ \midrule
                &  \multicolumn{4}{c}{1 energy level} && \multicolumn{4}{c}{8 energy levels}\\
                   \cmidrule{2-5}                     \cmidrule{7-10}
Image size      &   $256^2$ & $512^2$ & $1024^2$ & $2048^2$ && $256^2$ & $512^2$ & $1024^2$ & $2048^2$ \\
FoV [arcmin]    &    17 & 34&  68 & 136          &&  17 & 34&  68 & 136          \\
\cmidrule{2-10}
BippSsCpu       &   1.55&   1.39&   1.41&   1.70  &&   1.35&   1.42&   1.49&   1.68 \\
BippSsGpu       &   9.28&  11.01&  13.54&  17.58  &&   8.89&  11.77&  13.88&  17.42 \\
BippNufftCpu    &   1.64&   3.95&  13.40&  45.27  &&   0.49&   1.27&   4.27&  13.29 \\
BippNufftGpu    &   8.04&  24.97&  77.91& 207.72  &&   2.55&  10.05&  39.86& 114.06 \\
\cmidrule{2-10}
CASA            &   5.00&   8.37&  52.93& 133.49  &&   2.43&  10.08&  31.37& 106.78 \\
WSClean         &   4.42&  13.27&  49.57& 230.47  &&   3.94&  14.06&  50.21& 222.03 \\
\bottomrule
\end{tabular}
\caption{Speedup factors for \texttt{BIPP}, \texttt{CASA} and \texttt{WSClean} solutions compared to the reference Bluebild Python implementation to produce images with 256, 512, 1024 and 2048 pixels resolutions. BIPP speedups in the left part of the table were obtained, setting up 1 positive energy level, while on the right-hand side are given the speedup factors obtained when setting up 8 positive energy levels. The small differences observed for \texttt{CASA} and \texttt{WSClean} results between the 1 and 8 energy levels setups stem from instabilities inherent to using shared resources (\texttt{CASA} and \texttt{WSClean} were run by default along each solution).}
\label{tab:val:dirty_maps}
\end{table}

\begin{table}\centering
\setlength\extrarowheight{-3pt}
\begin{tabular*}{\linewidth}{@{\extracolsep{\fill}}lrr@{}}\toprule
Process                                &   Time [s]  &   \% Total   \\ \midrule
\textbf{Total}                        &  45.57    &  100.0 \\
\textbf{\textit{Parameter estimation}}  &   4.42    &    9.7 \\
\hspace{6pt}  \small Reading visibilities  &  \small 1.91    &   \small 4.2 \\
\hspace{6pt}  \small Processing            &  \small 2.50    &   \small 5.9  \\
\hspace{12pt} \small Eigen decomposition  &  \small 0.60    &   \small 1.3 \\
\textbf{\textit{Imaging}}               &  41.03    &  90.0 \\
\hspace{6pt} \small Reading visibilities         &   \small 1.84    &   \small 4.0\\
\hspace{6pt} \small Processing                   &  \small 39.19    &  \small 86.0 \\
\hspace{12pt} \small Eigen decomposition         &   \small 0.50    &   \small 1.1 \\
\textbf{\textit{Other}}                 &  0.12    &   0.3  \\
\bottomrule
\end{tabular*}
\caption{ Decomposition of the times to solution for the \texttt{BIPP} NUFFT GPU solution obtained with a single positive energy level when generating a 2048 x 2048 pixel image from 512 stations. Parameter estimation is an optional preliminary imaging step that runs on a small subset of the input file before the full image synthesis.}
\label{tab:perf:tts}
\end{table}

\section{Scientific Applications}
\label{sec:SA}

One of the unique aspects of \texttt{BIPP} is that the sky is reconstructed in distinct orthonormal eigenimages:
\begin{equation}
    \widetilde {I} = \sum_{a} \lambda_{a} |\epsilon_{a}|^2 =  \sum_{a} \lambda_{a} |\Psi \alpha_a|^2 ~.
    \label{eq:eigenvaluedeco}
\end{equation}
{ Each eigenimage} $ |\epsilon_{a}|^2 $ can be reconstructed from { the corresponding} visibility eigenvector $\alpha_a$ in parallel.
{ As discussed in Section~\ref{subsec:energy_levels_and_partitioning}, these} eigenimages and eigenvisibilities can be sorted and clustered by their eigenvalues $\lambda_a$, allowing for natural { partitioning of energy in the radio sky} into separate images.

This source separation is a unique benefit of the algorithm, and we present a few examples of this energy separation { on real and simulated data. These examples use $k$-means clustering to cluster distinct $\lambda_a$ into $k$ energy levels. }

\subsection{Point sources}\label{sec:61}

\begin{figure*}
    \centering
    \includegraphics[width=\textwidth]{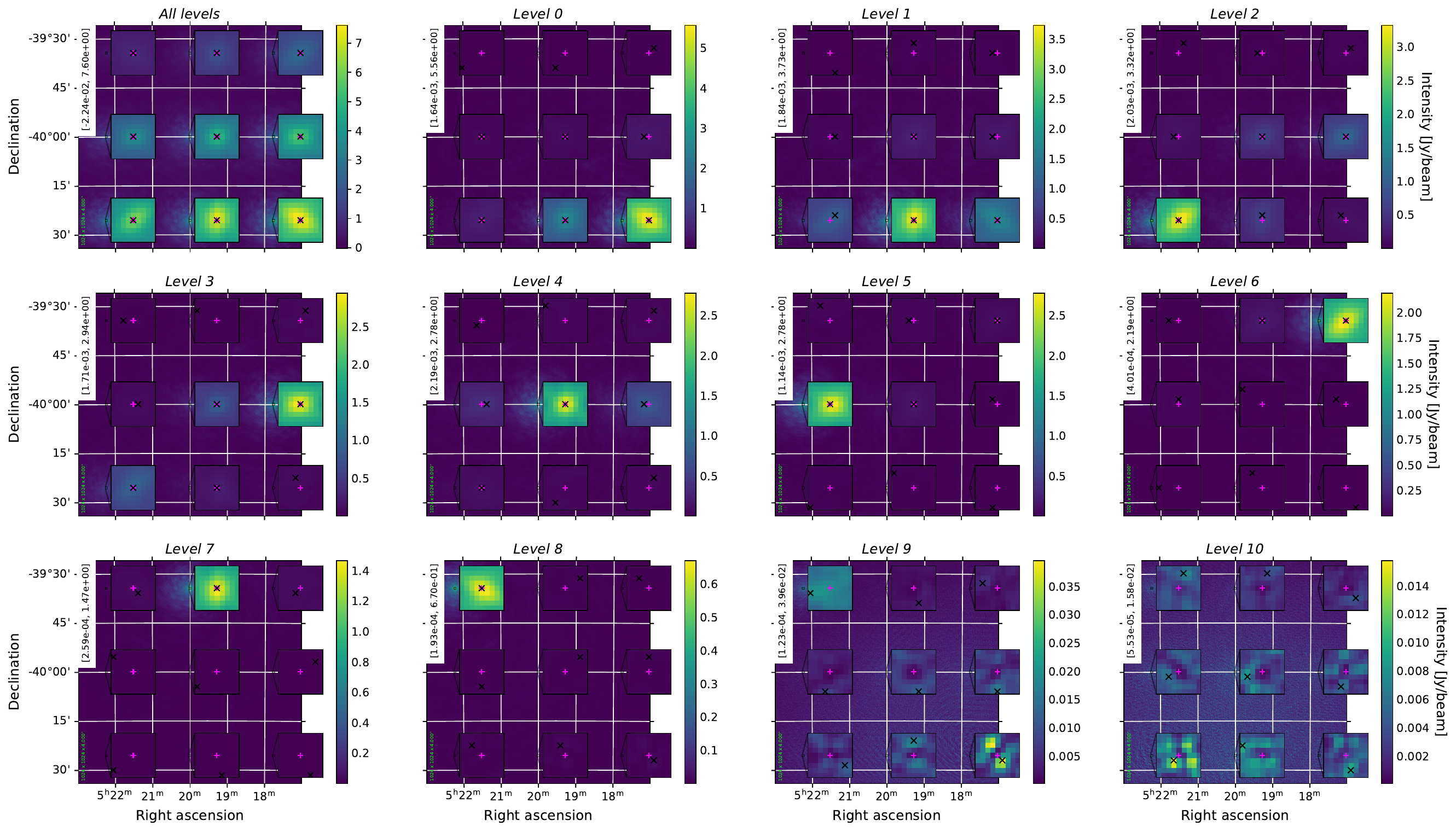}
    \caption{Simulated observation of 9 point sources with an SKA-Low configuration. The top left panel shows the LSQ image, and the other panels show the energy level partitioning in order of decreasing energy.}
    \label{fig:sa:gleam}
\end{figure*}

{ 
In Figure~\ref{fig:sa:gleam}, we show an example of the energy level partitioning performed by \BIPP on simulated point sources. We use the same SKA-Low observation configuration as described in Section~\ref{sec:valdata}, but we modify the sky model to have evenly spaced source flux of 1 Jy - 9 Jy. To primary beam correction is applied.

The reconstructed LSQ image and corresponding eigenvalue partitioning are shown in Figure~\ref{fig:sa:gleam}. The LSQ image is shown in the upper left panel. Although the bottom left source has a true flux of 9 Jy, due to the primary beam it is reconstructed with a flux of $\sim 7$ Jy/beam. The energy partitioning creates distinct images of flux at different scales but there is some mixing of components between different levels-- we can see for example that the brightest source is split into a $\sim 5.5$ Jy/beam source in Level 0 and a $\sim 1.5$ Jy/beam source in Level 1. Some sources are almost perfectly partitioned into their own levels, as seen in Levels 5, 6, 7, and 8. Levels 9 and above begin to contain artifacts consistent with noise.
}

\subsection{Solar Limb Brightening} \label{sec:62}
\begin{figure*}
\centering
\begin{subfigure}{.48\linewidth}
  \centering
  \includegraphics[width=0.99\linewidth]{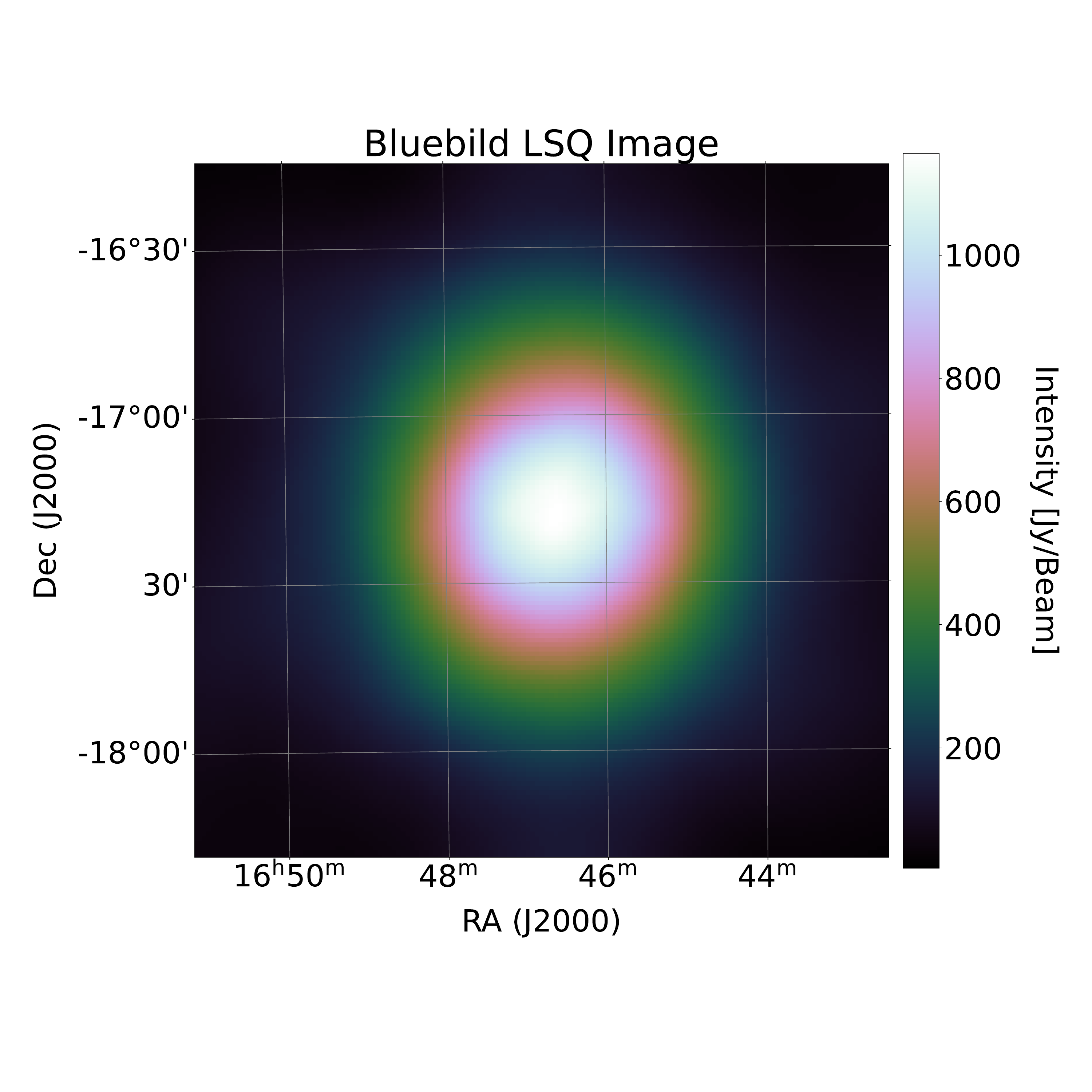}
  \caption{}
\end{subfigure}%
\begin{subfigure}{.48\linewidth}
  \centering
  \includegraphics[width=0.8\linewidth]{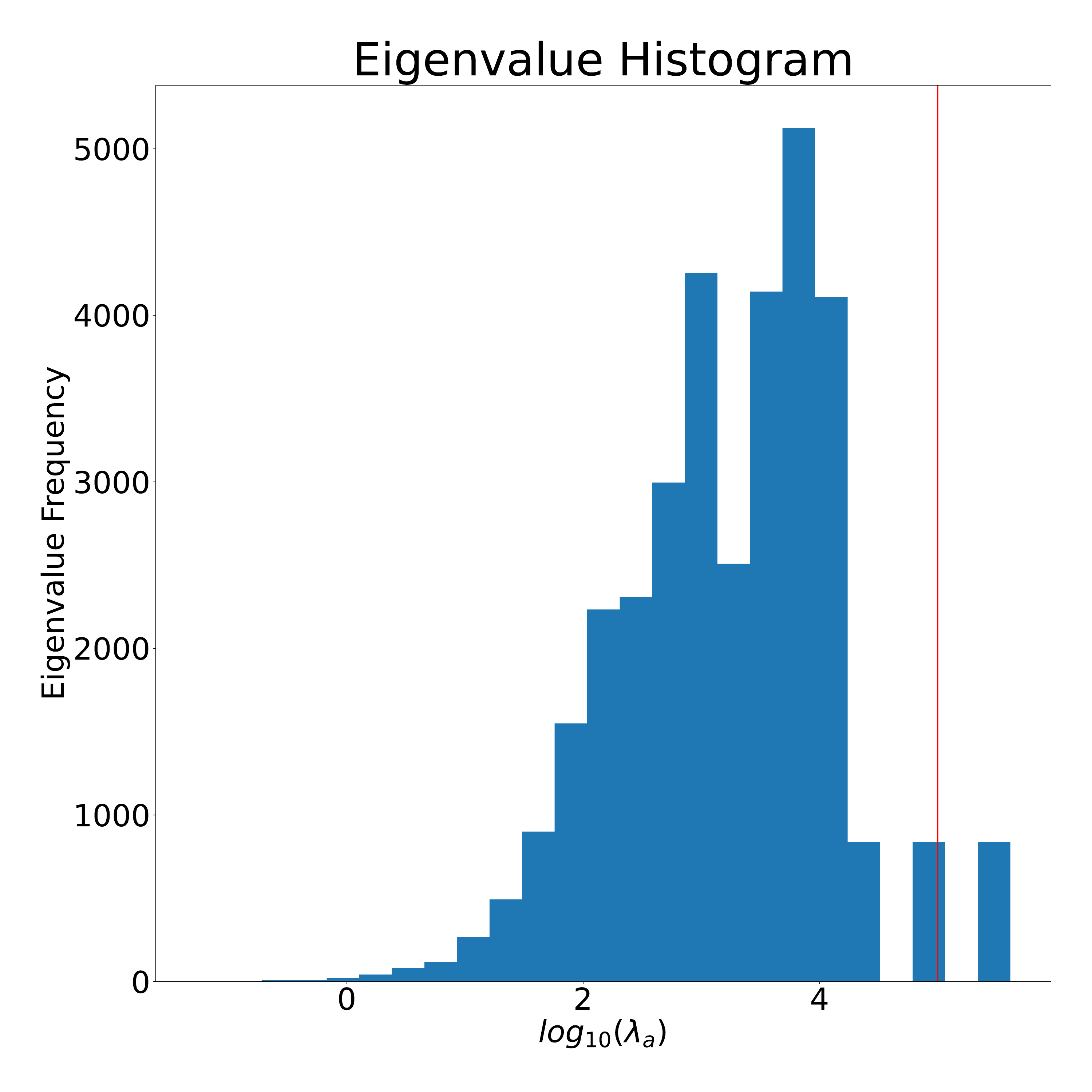}\vspace{3em}
  \caption{}
\end{subfigure}
\begin{subfigure}{.48\linewidth}
  \centering
  \includegraphics[width=0.99\linewidth]{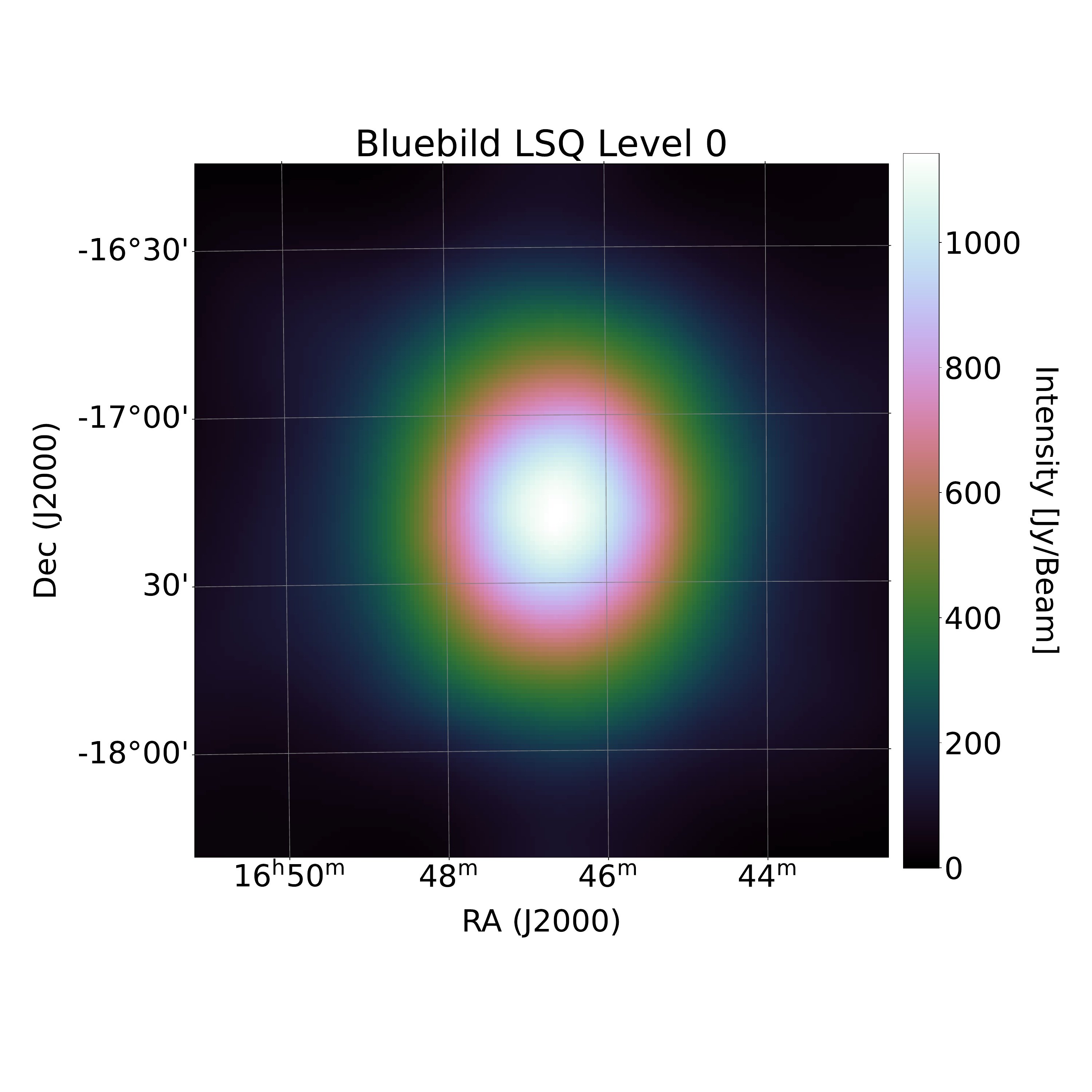}
  \caption{}
\end{subfigure}
\begin{subfigure}{.48\linewidth}
  \centering
  \includegraphics[width=0.99\linewidth]{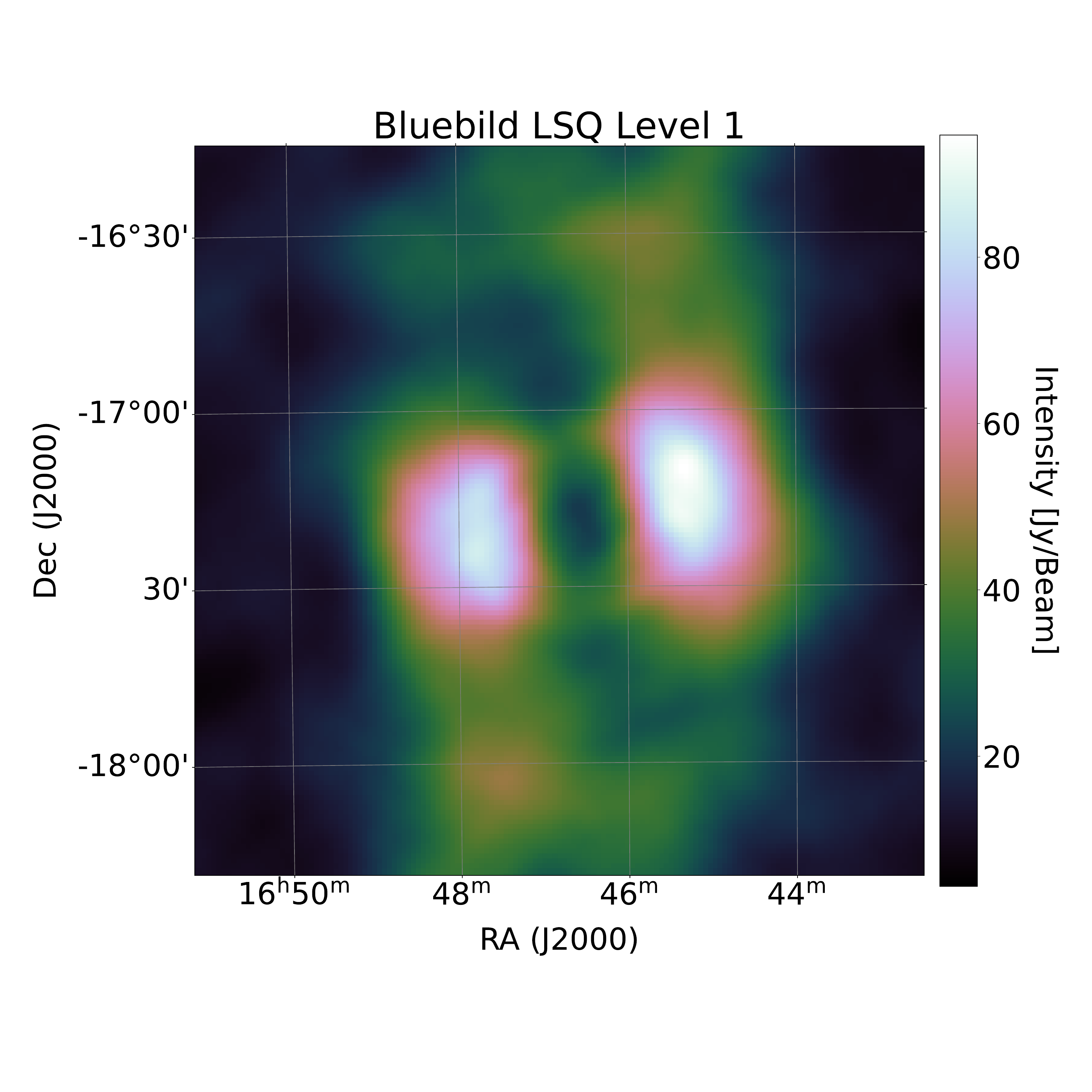}
  \caption{}
\end{subfigure}%
\caption{MWA Observation of the quiet sun decomposed into two eigen-levels by \texttt{BIPP}'s functional PCA. (a) shows the least-squares image obtained by summing all least-squared eigen-levels. (b) shows the eigenvalue histogram corresponding to this observation. The red line within the top-right panel shows where the k-means clustering separates least-squared level 0 and least-squared level 1. (c) shows the least-squared level 0 of the MWA solar observation, while (d) shows the least-squared level 1 of the MWA solar observation. Of particular note is the solar limb brightening emission seen in (d), which is not accessible in the dirty images produced by CASA and WSClean.}
    \label{fig:sa:solar_observation}
\end{figure*}

In Figure \ref{fig:sa:solar_observation} we show an example of the source separation performed by \texttt{BIPP} on observed MWA data. The observation targets the quiet sun \citep{Rohit_Paper} using Phase I of the MWA; meaning that 128 phased arrays observe free-free emission from the solar corona which is affected by anisotropic scattering and refraction due to the coronal medium. The resulting image has a field of view of 14.3 degrees. We zoom in on a 2.1 degree field of view centered on the sun in panels (a), (c) and (d) of Figure \ref{fig:sa:solar_observation}. The integration time for the observation is 10 seconds, with 0.5 second time steps. The central frequency of observation is 216.94 MHz; further, the observation is composed of 64 channels of bandwidth $\Delta \nu$ = 40 kHz between 215.68 MHz and 218.2 MHz.

We image this observation with \BIPP, shown in  Figure \ref{fig:sa:solar_observation}.  We note the radially decreasing emission that is typical of radio images of the sun in Figure \ref{fig:sa:solar_observation} (a). 
We reconstruct the image in two separate levels. The higher level (level 0; Figure \ref{fig:sa:solar_observation} (c)) is almost identical to the summed least-squared image in the top-left panel of the same figure. However, the lower level (level 1; Figure \ref{fig:sa:solar_observation} (d)) shows an interesting emission pattern focused on the solar limb. This emission is not evident in images reconstructed with \texttt{CASA} or \texttt{WSClean}, but is revealed through Bluebild's eigenvalue decomposition.

\begin{figure*}
\centering
\begin{subfigure}{.33\linewidth}
  \centering
  \includegraphics[width=0.9\linewidth]{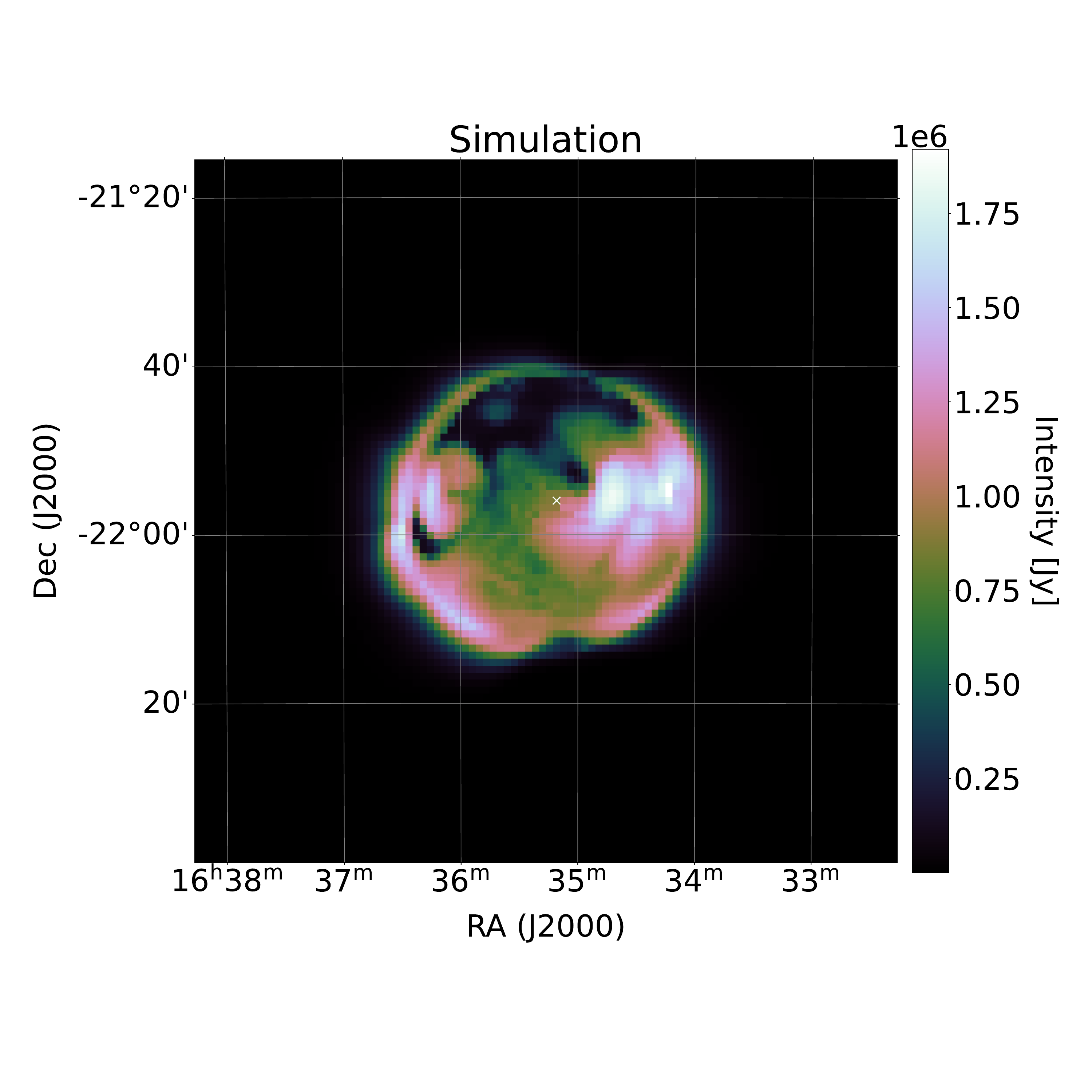}
  \caption{}
\end{subfigure}%
\begin{subfigure}{.33\linewidth}
  \centering
  \includegraphics[width=0.9\linewidth]{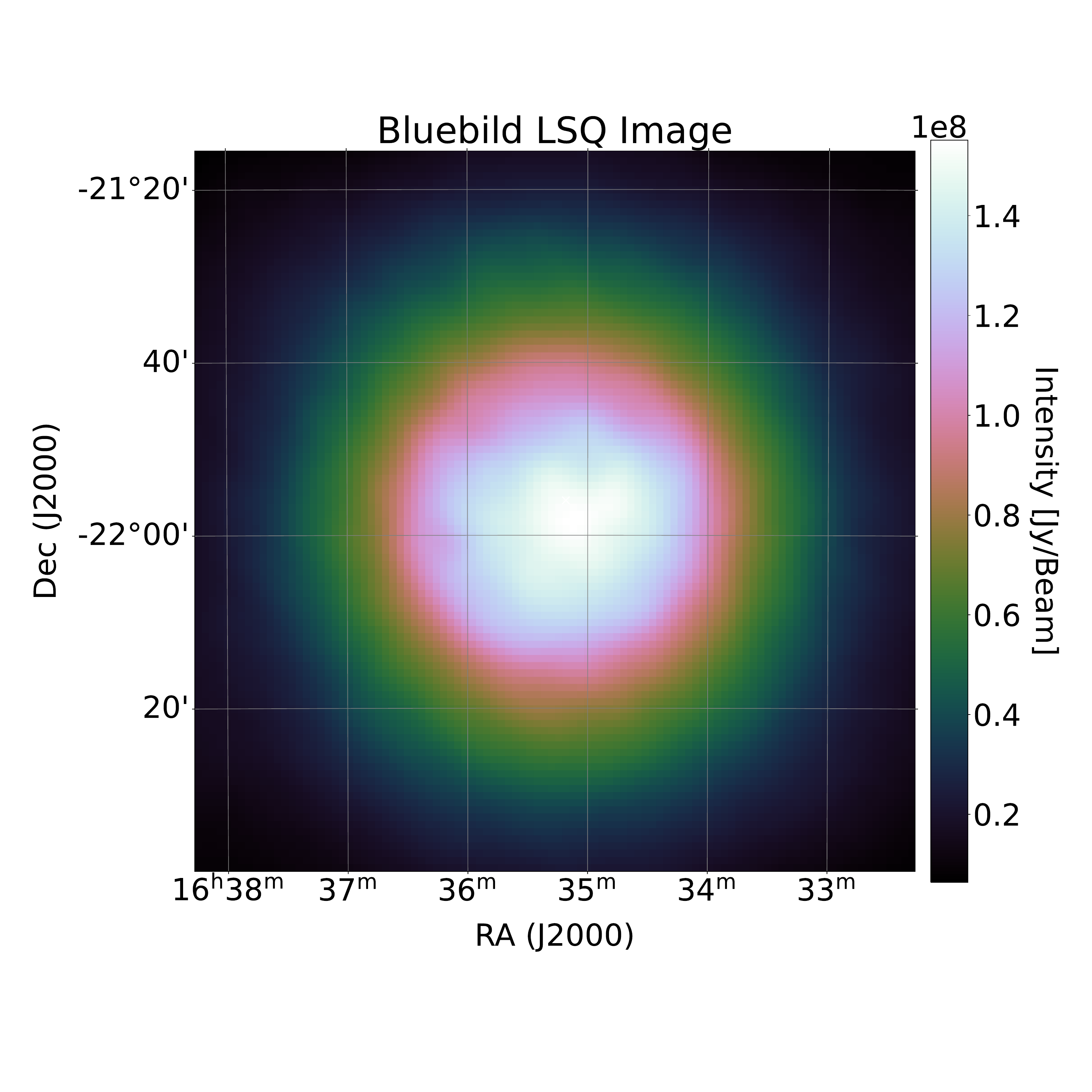}
  \caption{}
\end{subfigure}
\begin{subfigure}{.33\linewidth}
  \centering
  \includegraphics[width=0.75\linewidth]{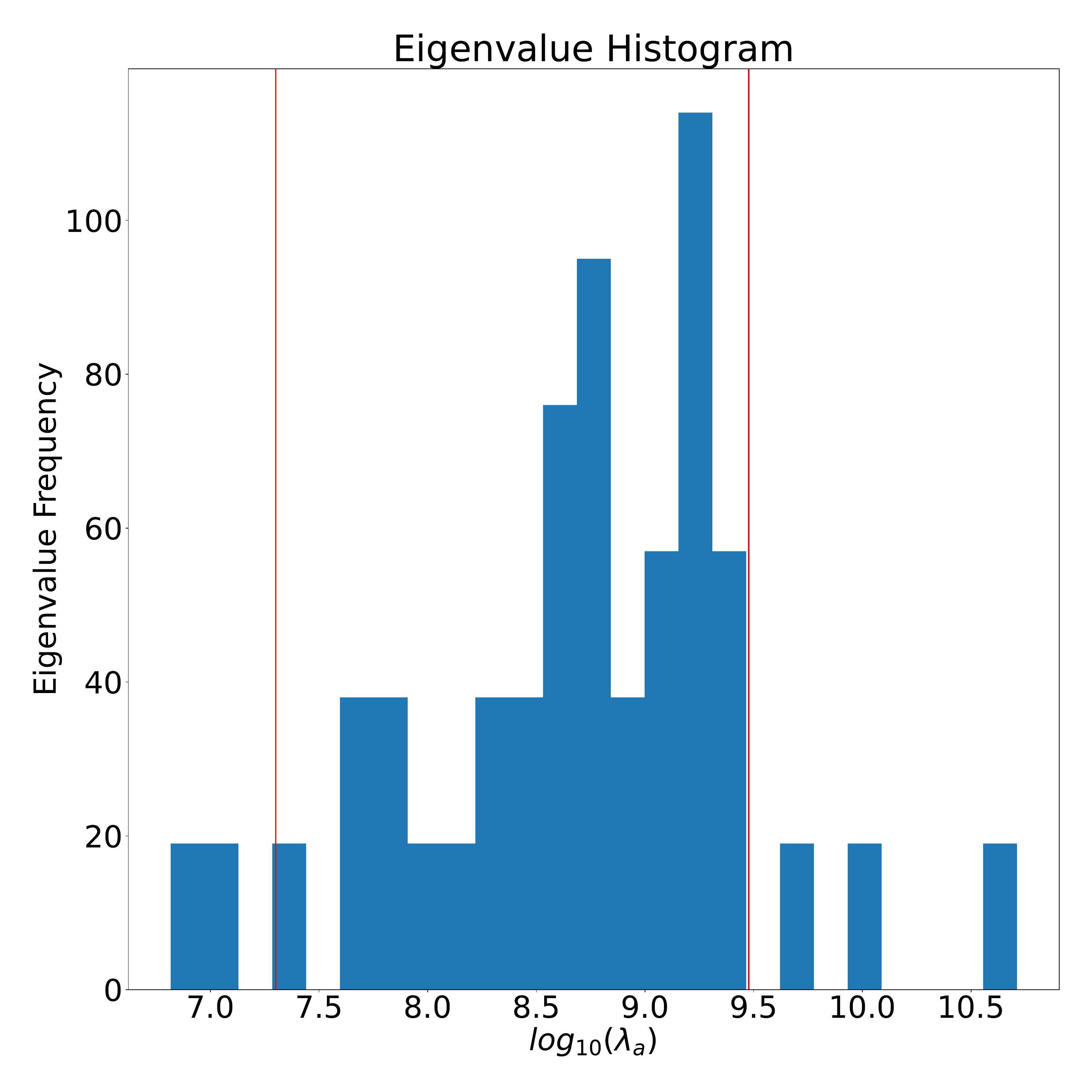}
  \vspace{1.5em}
  \caption{}
\end{subfigure}
\begin{subfigure}{.33\linewidth}
  \centering
  \includegraphics[width=0.9\linewidth]{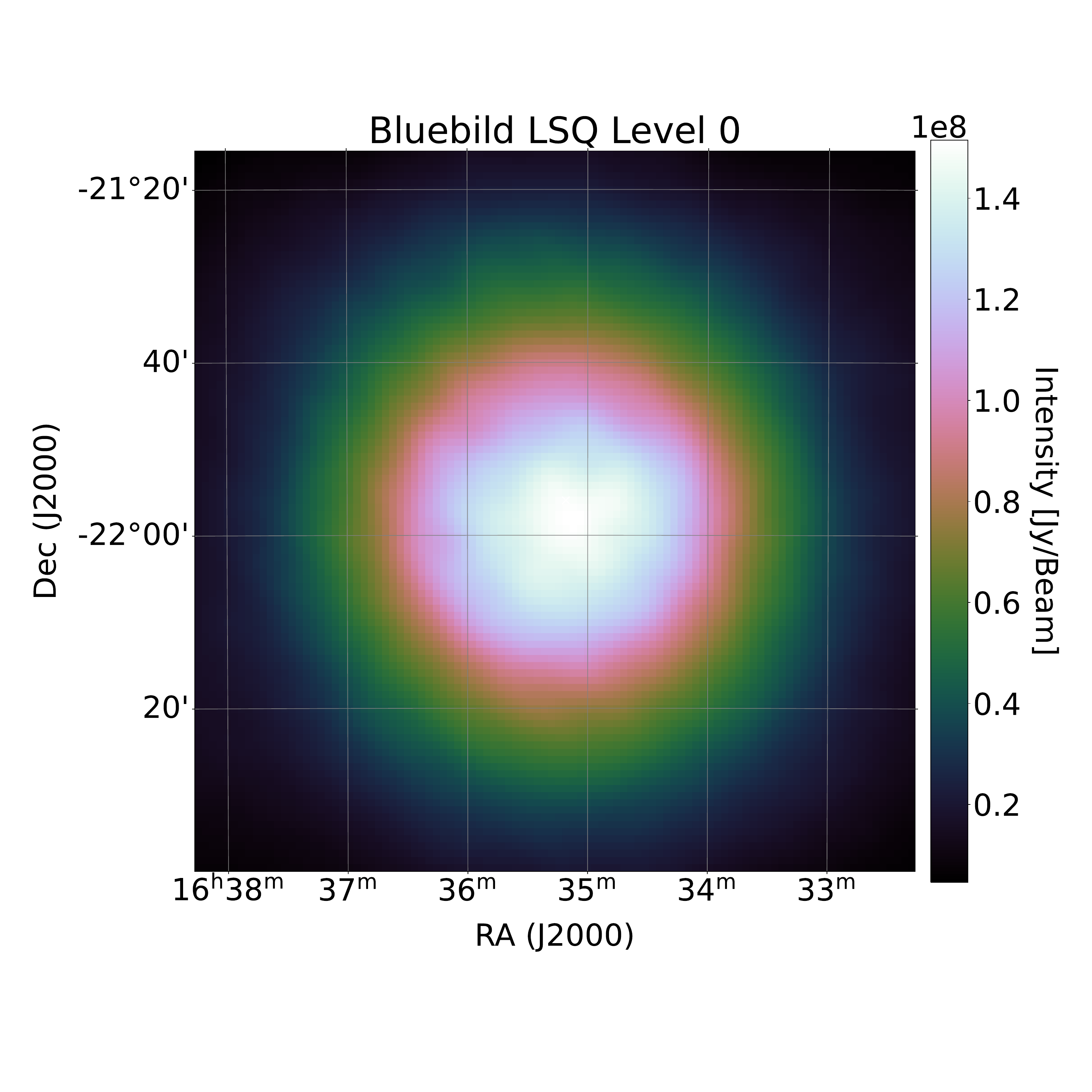}
  \caption{}
\end{subfigure}%
\begin{subfigure}{.33\linewidth}
  \centering
  \includegraphics[width=0.9\linewidth]{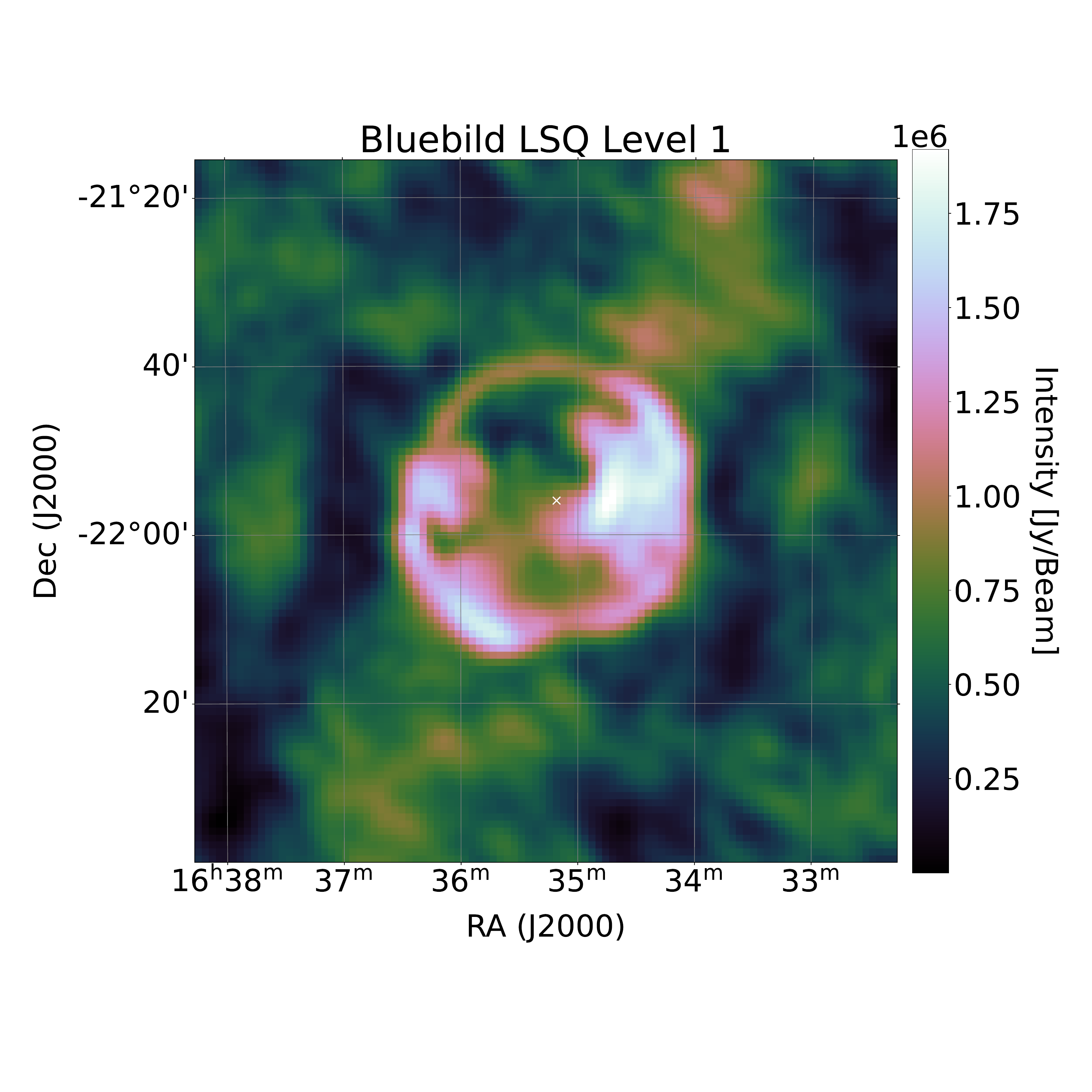}
  \caption{}
\end{subfigure}
\begin{subfigure}{.33\linewidth}
  \centering
  \includegraphics[width=0.9\linewidth]{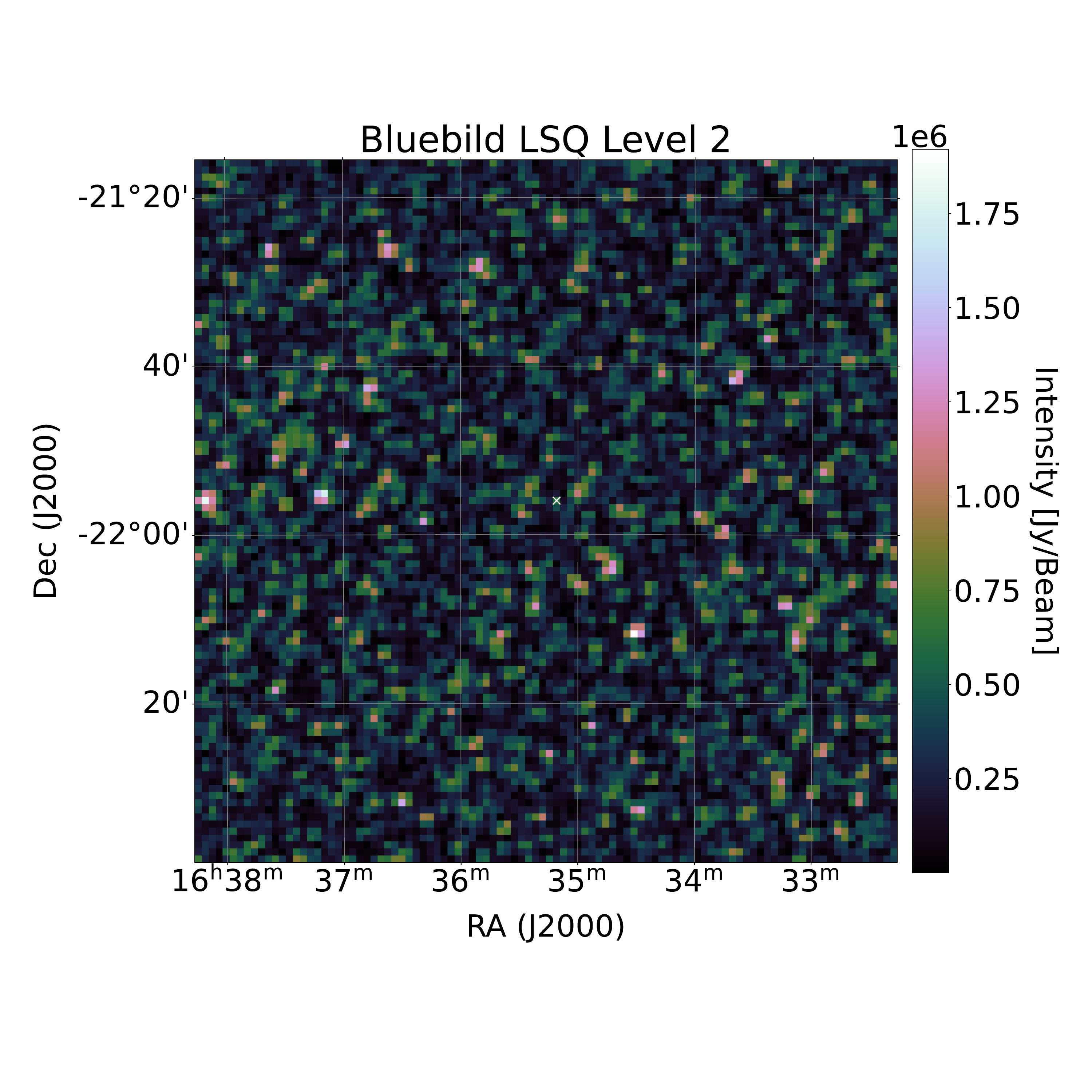}
  \caption{}
\end{subfigure}
\caption{MWA Mock-Observation of \texttt{FORWARD} Solar Simulation. (a) shows the \texttt{FORWARD} simulation that is used as an input sky model to \texttt{OSKAR}. (b) shows the summed \texttt{BIPP} least-squared image. (c) shows the eigenvalue histogram corresponding to this mock-observation, with the red lines showing the separation between different eigen-levels. The bottom row, (d),(e) and (f) shows the least-squares levels 0, 1 and 2. Note the presence of finer features in least-squares level 1, similar to those seen in the \texttt{FORWARD} simulation. Also noteworthy is the absence of these features from (b) which is consistent with the CASA and WSClean dirty images.}
    \label{fig:sa:solar_simulation}
\end{figure*}

We image a simulated solar observation (Figure \ref{fig:sa:solar_observation} (a)) using \texttt{OSKAR} to validate our discovery of radio limb brightening emission from the sun, shown in Figure~\ref{fig:sa:solar_simulation}.

We simulate the solar free-free emission using the \texttt{FORWARD} \citep{Forward} software. This software uses a self-consistent Magnetohydrodynamic Algorithm outside a Sphere (MAS) coronal model. It then takes the temperature, electron density and magnetic field from input HMI Magnetograms and normalises these against photospheric values. Finally, it calculates the brightness temperature in various Stokes parameters. We calculate the corresponding visibilites using the MWA Phase I configuration with \texttt{OSKAR}.

Figure \ref{fig:sa:solar_simulation} (b) shows the summed \texttt{BIPP} least-squared image. We can see the radially decreasing, centre brightened solar emission, but not the fine structure present in the simulation itself.

However, we can observe this fine scale structure in the middle imaging level (level 1; Figure \ref{fig:sa:solar_simulation} (e)). Like the simulation, emission is primarily situated around the solar limb and is roughly two orders of magnitude weaker than the radially decreasing emission seen in the summed \texttt{BIPP} least-squares image. The presence of such features in the least-squared level 1 validates our observation of solar limb-brightening in the bottom-right panel of Figure \ref{fig:sa:solar_observation}. Finally the Figure \ref{fig:sa:solar_simulation} (f) shows { the noise} in level 2.

This example is particularly interesting, since it demonstrates both the denoising and the filtering properties of \texttt{BIPP}'s eigenvalue decomposition and k-means clustering.

{ 
\subsection{Discussion}
As discussed in Section~\ref{sec:vsclean}, the Bluebild LSQ image $\widetilde I$ does not include any regularization, whereas the final image created by clean $I_{k \rightarrow \infty}$ satisfies the LSQ problem and includes regularization.
If we wish to perform deconvolution on the output of \bipp, we can set $G_\Psi=I$ in the eigenvalue decomposition in Equation~\ref{eq:eigenvaleq} to solve for the eigenpairs $\{ \lambda'_a, \alpha_a'\}$:
\begin{equation}
V \alpha_a' = \lambda'_a \alpha_a'.
\end{equation}
This gives us an eigendecomposition of the dirty image $\hat I$ instead of $\widetilde I$:
\begin{align}
    \hat B &= \Psi V \Psi^* = \sum_a^{N_A} \lambda_a' \Psi \alpha_a' (\alpha_a')^H \Psi^* \\
    \hat I &=  \sum_a^{N_A} \lambda_a' \Psi |\alpha_a'|^2 \Psi^* 
\end{align}
where $\hat B_{jj} = \hat I_j$. 

Because for most interferometers $G_\Psi \approx I$, this is also a useful framework for interpreting the results of the decomposition in Sections~\ref{sec:61} and~\ref{sec:62}.
The fPCA gives us orthogonal visibility eigenvectors $\alpha_a'$ that can be linearly combined to create the dirty image $\hat I$. Each eigenvisibility $|\alpha_a'|^2$ should correspond to sources of energy or ``components'' corresponding to the eigenvalue $\lambda_a'$. This separation does not impose any prior on the morphology of the components, unlike CLEAN, but also does not apply any regularization so the different energy levels are still contaminated by the instrument PSF.

The eigenvalue decomposition is particularly useful for scientific applications that need to filter specific energy levels. Bright source peeling can be performed in visibility space by discarding the brightest eigenvisibilities $|\alpha|^2$ before the image synthesis step.

The energy levels produced by \texttt{BIPP} could also be a useful input to AI-based deconvolution, as the dynamic range of each energy level is much smaller than the combined LSQ image. Regularization/cleaning could be performed independently and in parallel on each level, either in visibility space or image space, before combining the levels to create the cleaned LSQ image.

However, the decomposition offered by Bluebild is not always easily interpretable. The energy $\lambda_a'$ does not necessarily correspond to the energy of a particular component. For example, in Figure~\ref{fig:sa:gleam} several point sources are split into multiple components in the same location across different energy levels.
}

\section{Conclusions}

We have presented \texttt{BIPP}, an HPC implementation of the Bluebild algorithm. \BIPP offers an alternative strategy for interferometric imaging, producing LSQ estimates of the sky using fPCA to reconstruct the sky in distinct energy levels and levering the {  3D NUFFT to image on the sphere}.

{ Using \BIPP and the new NUFFT Synthesis algorithm we are able to use the Bluebild algorithm with SKA-like array configurations and large image sizes. We show that \texttt{BIPP}} reconstructs observations with { comparable} image fidelity compared to the \wsclean~$w$-stacking implementation and improved time-to-solution{  for image sizes with $N_\text{pixel} < 1024^2$.} While \texttt{BIPP} does not perform any regularization and the PSF must be removed as a post-processing step, it is a useful option for deconvolution-free imaging.

{ We see that for interferometers such as LOFAR, MWA, MeerKAT, and the SKA Mid and Low configurations, the Gram matrix correction has a small effect and the Bluebild LSQ image $\widetilde I$ is very similar to the dirty image $\hat I$. The Gram matrix correction has a stronger effect if the baseline separation is equal to or smaller than the observing wavelength, which is true for station-level elements of phased array interferometers. However, beamforming is usually performed at the station level and the individual station element data are not available. It is possible to leverage the eigendecomposition offered by Bluebild on the dirty image $\hat I$, which would allow  CLEAN algorithms to be run in parallel on the distinct energy levels.  }

 The eigendecomposition provided by \texttt{BIPP} may also allow for scientific analysis that is not possible with combined images. For example, noise suppression can be obtained by discarding the lowest eigenlevels,  foreground removal can be achieved by discarding the highest eigenlevels, and interesting substructure may be reveals in intermediate eigenlevels, as shown in Figure~\ref{fig:sa:solar_simulation}. { Additonally, the decomposition reduces the dynamic range of images, which could be a useful input to AI-based cleaning or analysis techniques.}

Future HPC developments of \BIPP will focus on scalability, node-level parallelism of the NUFFT, and optimizing our GPU implementation of the type-3 nufft for radio astronomy data. { We will also investigate integrating the WStackingGridder\footnote{\href{http://www.andreoffringa.org/wsclean/doxygen/classWStackingGridder.html}{http://www.andreoffringa.org/wsclean/doxygen/classWStackingGridder.html}} as an alternative image synthesis option.}

\section*{Acknowledgements}

This work was supported by the  Platform for Advanced Scientific Computing (PASC) project ``Next-Generation Radio Interferometry.'' SK acknowledges the financial support from the SNSF under the Sinergia Astrosignals grant (CRSII5\_193826). This work has been done in partnership with the SKACH consortium through funding by SERI, and was supported by EPFL through the use of the facilities of its Scientific IT and Application Support Center (SCITAS). The authors gratefully acknowledge the use of facilities of the Swiss National Supercomputing Centre (CSCS). Development of the Bluebild algorithm was supported via the ASTRON-IBM Dome project by the Dutch Ministry of Economic Affairs and by the Province of Drenthe.

We would also like to thank Prof. Oleg Smirnov, Landman Bester, Jonathan Kenyon, and Simon Perkins (SARAO/Rhodes University) for helpful discussions and suggestions regarding interferometry and calibration.

\section*{Data Availability}

Most of the datasets used in validation are simulated observations generated by open-source libraries as documented in the text. These simulated datasets can be made available upon reasonable request to the authors. 
The real LOFAR data used for validation are available at DOI:\href{https://doi.org/10.5281/zenodo.1042525}{10.5281/zenodo.1042525}.
 The \BIPP source code is publically available on GitHub\footnote{\href{https://github.com/epfl-radio-astro/bipp}{https://github.com/epfl-radio-astro/bipp}}.


\bibliographystyle{elsarticle-harv} 
\bibliography{refs} 




\appendix

\section{Effect of the Gram Matrix}
\label{sec:appGram}

\begin{figure}
    \centering
    \includegraphics[width=0.4\textwidth]{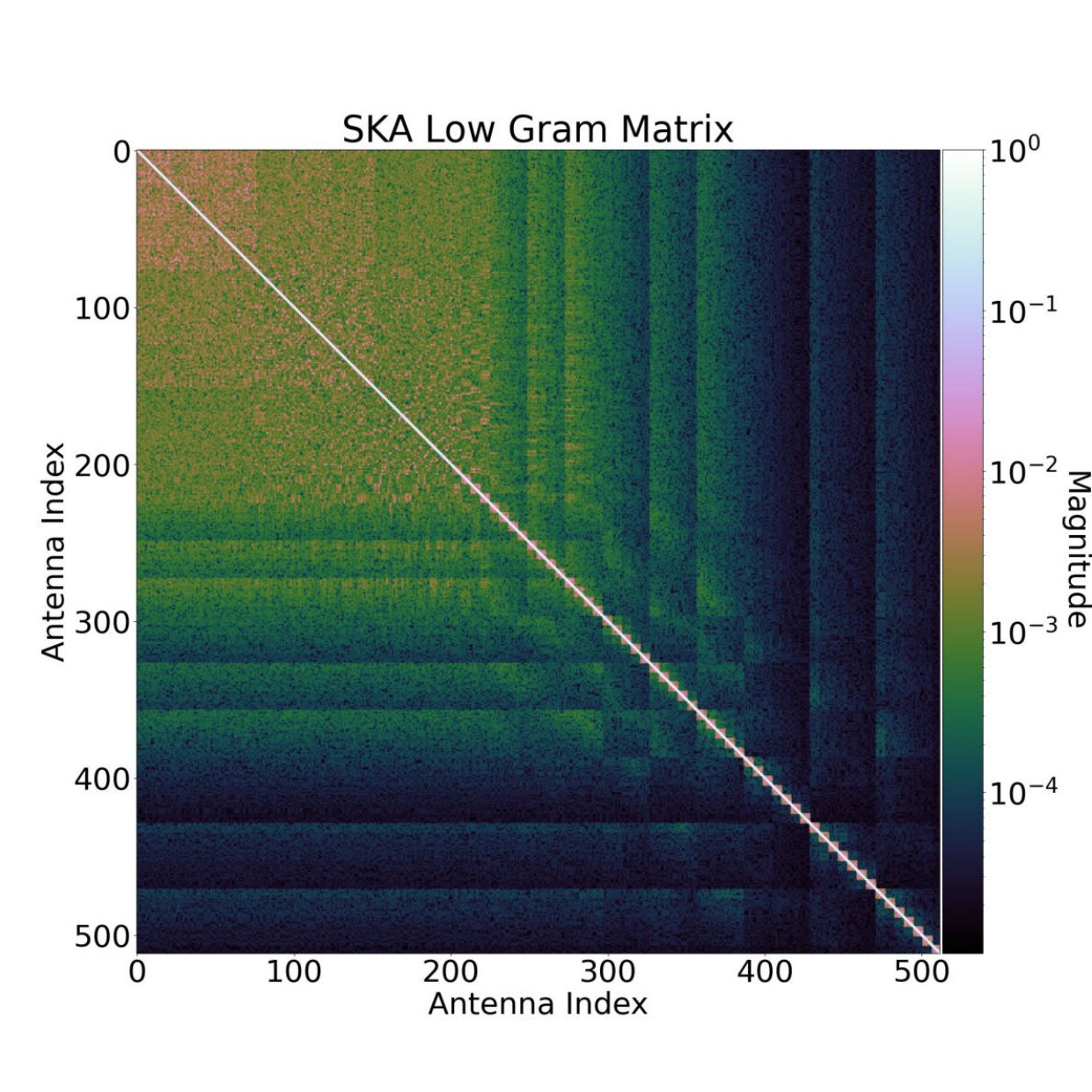}
    \caption{The Gram matrix visualised for the simulated SKA-Low array at 53 MHz.}
    \label{fig:appendix:Gram:GramMatrix}
\end{figure}

{ 
Recall that the eigenvalue-eigenvector pairs $\{ \lambda_a, \alpha_a\}$ are found by solving the generalized eigenvalue problem $V\alpha_a = \lambda_a G_\Psi \alpha_a$, therefore:
\begin{equation}
    V = G_\Psi ~ A ~ \Lambda~ A^H~ = G_\Psi V',
\end{equation}
where the columns of matrix $A \in \mathbb{C}^{N_A \times N_A}$ are the ordered eigenvectors $\alpha_a$, $\Lambda$ is a diagonal matrix with diagonal elements as the ordered eigenvalues $\lambda_a$, and V' is what we call the Gram-corrected visibilities.

The least-squares consistent image $\widetilde I$ constructed by Bluebild is given by:
\begin{align}
\widetilde I &= \Psi V' \Psi^* \\
\widetilde I_j &= \sum_n^{N_A \times N_A} V_n' e^{i \langle \vec b_n,  \vec r_j \rangle}~,
\end{align}
and the back-projected image $\hat I$ is given by:
\begin{align}
\hat I &= \Psi V \Psi^* = \Psi G_\Psi V' \Psi^* \\
\hat I_j &= \sum_n^{N_A \times N_A} (G_\Psi V')_n ~ e^{i \langle \vec b_n,  \vec r_j \rangle}~,
\end{align}

In the context of an NUFFT, the two images have the same input coordinates $\vec b_n$ and output coordinates $\vec r_j$, but the sample values differ. We can use the continuous expression of the Gram matrix from Equation~\ref{eq:gram} to expand our expression for $V$ in terms of $V'$:
\begin{align}
    V_{pq} &= \sum_k^{N_A} (G_\Psi)_{pk} V_{kq}' \\
    &= \sum_k^{N_A} \mathrm{sinc}~\Bigl(2~ \Bigl|\Bigl| \frac{ \vec b_{pk} f }{c} \Bigr|\Bigr| \Bigr) \ V_{kq}' \\
     &= V_{pq}' +  \sum_{k \neq p}^{N_A} \mathrm{sinc}~\Bigl(2~ \Bigl|\Bigl| \frac{ \vec b_{pk}  }{\lambda} \Bigr|\Bigr| \Bigr) \ V_{kq}' 
\end{align}
using $\lambda = f/c$.  Each visibility $V_{pq}$ can be expressed as a linear combination of the Gram-corrected visibilities $V'$. The effect of this mixing is larger sidelobes in $\hat I$ compared to $\widetilde I$. 

Note that $\text{sinc}(x) = \sin(\pi x)/ \pi x$, so if  $||\vec b_{pk}|| \gg \lambda$ for all $k \neq p$ then $V_{pq} \simeq V_{pq}'$. This is the case for most modern radio interferometers, so the difference between $\hat I$ and $\widetilde I$ is tiny.

To illustrate the difference between $\hat I$ and $\widetilde I$,
}
using \texttt{OSKAR} we create simulated SKA-Low telescope observations with
a configuration of 512 stations and image a sky model with a single 1000 Jy source. We image at 53MHz and 89MHz as the off-diagonal terms of the Gram matrix become larger at lower frequencies as shown in Eq.~\ref{eq:gram}. When imaging at 53MHz we use a FoV of 1.16692 degrees and when imaging at 89MHz we use a FoV is is 0.7002 degrees. This sky model is observed over 1 epoch. The resulting Gram matrix for 53MHz can be seen in Figure \ref{fig:appendix:Gram:GramMatrix}, and contains { nonzero} off-diagonal terms.
 
The visibilities obtained from these simulated observations are then imaged using \texttt{BIPP} with and without using the Gram matrix and shown Figures \ref{fig:appendix:Gram:GramMatrixComparison} and \ref{fig:appendix:Gram:GramMatrixHighFrequency}. The presence of negative artefacts from the interferometric sidelobes in the residual image tells us that the sidelobes are more intense when using BIPP without the Gram matrix compared to using BIPP with the Gram matrix. Furthermore, we see in Figure \ref{fig:appendix:Gram:GramMatrixHighFrequency} that this effect is more pronounced at lower frequencies. This can be observed in the magnitude of the residuals, which are stronger for the 53 MHz case compared to the 89 MHz case.

\begin{figure*}
    \centering
    \includegraphics[width=0.99\textwidth]{fig_BIPP_Paper_Appendix_Figure_A11-compressed.pdf}
    \caption{The effect of using the Gram matrix for the generalized eigenvalue decompositon in BIPP reconstructed images at 53MHz. The left column subplot contains the BIPP image normalised by its maximum value. The middle subplot contains the reconstructed BIPP image with the Gram matrix set to identity. This has also been normalised by its maximum value. The right subplot shows the residual obtained when taking the difference between the BIPP images reconstructed with and without the Gram matrix. The presence of negative (blue) features in the residual image tells us that the sidelobes are slightly stronger when the Gram matrix is turned off.}
    \label{fig:appendix:Gram:GramMatrixComparison}
\end{figure*}

\begin{figure*}
    \centering
    \includegraphics[width=0.99\textwidth]{fig_BIPP_Paper_Appendix_Figure_A12-compressed.pdf}
    \caption{The effect of using the Gram matrix for the generalized eigenvalue decompositon in BIPP reconstructed images at 89MHz. The left column subplot contains the BIPP image normalised by its maximum value. The middle subplot contains the reconstructed BIPP image with the Gram matrix set to identity. This has also been normalised by its maximum value. The right subplot shows the residual obtained when taking the difference between the BIPP images reconstructed with and without the Gram matrix.The presence of negative (blue) features in the residual image tells us that the sidelobes are slightly stronger when the Gram matrix is turned off.}
    \label{fig:appendix:Gram:GramMatrixHighFrequency}
\end{figure*}


\label{lastpage}
\end{document}